\documentclass{aa}  

\usepackage{graphicx}
\usepackage{multirow}
\usepackage{colortbl}
\usepackage[table,x11names]{xcolor}
\usepackage{txfonts}

\begin{document} 

  \title{A first estimate of the Milky Way dark matter halo spin}

   \author{Aura Obreja\inst{1}\fnmsep\thanks{obreja@usm.lmu.de},
        Tobias Buck\inst{2}
        \and
        Andrea V. Macci\`{o}\inst{3,4,5}}

   \institute{Universit\"{a}ts-Sternwarte M\"{u}nchen, Scheinerstra\ss e 1, D-81679 M\"{u}nchen, Germany\\
   \and 
    Leibniz-Institut f\"{u}r Astrophysik Potsdam (AIP), An der Sternwarte 16, D-14482 Potsdam, Germany\\
             \and 
             New York University Abu Dhabi, PO Box 129188, Saadiyat Island, Abu Dhabi, United Arab Emirates\\
             \and 
             Center for Astro, Particle and Planetary Physics (CAP$^3$), New York University Abu Dhabi\\
             \and
            Max-Planck-Institut f\"{u}r Astronomie, K\"{o}nigstuhl 17, D-69117 Heidelberg, Germany}

   \date{Received April 1, 2021; accepted September 27, 2021}

  \abstract
  {The spin, or normalized angular momentum $\lambda,$ of dark matter halos in cosmological simulations follows a log normal distribution and has little correlation with galaxy observables such as stellar masses or sizes. There is currently no way to infer the $\lambda$ parameter of individual halos hosting observed galaxies. Here, we present a first attempt to measure $\lambda$ starting
  from the dynamically distinct disks and stellar halos identified in high-resolution cosmological simulations with the \texttt{Galactic Structure Finder (gsf)}. 
  In a subsample of NIHAO galaxies analyzed with \texttt{gsf}, 
  we find tight correlations between the total angular momentum of the dark matter halos, $J_{\rm h}$, and the azimuthal angular momentum, $J_{\rm z}$, of the dynamical distinct stellar components of the form: log($J_{\rm h}$)=$\alpha$+$\beta\cdot$log($J_{\rm z}$). The stellar halos have the tightest relation with $\alpha=9.50\pm0.42$ and $\beta=0.46\pm0.04$. The other tight relation is with the disks, for which $\alpha=6.15\pm0.92$ and $\beta=0.68\pm0.07$. While the angular momentum is difficult to estimate for stellar halos, there are various studies that calculated $J_{\rm z}$ for disks. In application to the observations, we used Gaia DR2 and APOGEE data to generate a combined kinematics-abundance space, where the Galaxy's thin and thick stellar disks stars can be neatly separated and their rotational velocity profiles, $v_{\rm\phi}(R),$ can be computed. 
  For both disks, $v_{\rm\phi}(R)$ decreases with radius with $\sim$2~km~s$^{\rm -1}$~kpc$^{\rm -1}$ for $R\gtrsim5$~kpc, resulting in velocities of $v_{\rm\phi,thin}=221.2\pm0.8$~km~s$^{\rm -1}$ and $v_{\rm\phi,thick}=188\pm3.4$~km~s$^{\rm -1}$ at the solar radius.
  We use our derived $v_{\rm\phi,thin}(R)$ and $v_{\rm\phi,thick}(R)$ together with the mass model for the Galaxy of \citet{Cautun:2020} to compute the angular momentum for the two disks: $J_{\rm z,thin}=(3.26\pm0.43)\times10^{\rm 13}$ and $J_{\rm z,thick}=(1.20\pm0.30)\times10^{\rm 13}$ M$_{\rm\odot}$kpc$\rm\cdot$km$\rm\cdot$s$^{\rm -1}$, where the dark halo is assumed to follow a contracted NFW profile. Adopting the correlation found in simulations, the total angular momentum of the Galaxy's dark halo is estimated to be $J_{\rm h}=2.69^{\rm +0.37}_{\rm -0.32}10^{\rm 15}$M$_{\rm\odot}$kpc$\rm\cdot$km$\rm\cdot$s$^{\rm -1}$ and the spin estimate is $\lambda_{\rm MW}=0.061^{\rm +0.022}_{\rm -0.016}$, which translates into a probability of 21\% using the universal log normal distribution function of $\lambda$. 
If the Galaxy's dark halo is assumed to follow a NFW profile instead, the spin becomes  $\lambda_{\rm MW}=0.088^{\rm +0.024}_{\rm -0.020}$, making the Milky Way a more extreme outlier (with a probability of only 0.2\%).} 
   

   \maketitle
%

\section{Introduction}

In the current mainstream model of structure formation, cold dark matter (DM) hierarchically clusters under the effect of gravity \citep{Peebles:1965} into the so-called cosmic web \citep{Bond1996} from the tiny density fluctuations present in the Universe at the epoch of recombination \citep[][]{Sachs:1967,Silk:1968,Peebles:1982}. In this scenario, gas cools and collapses into the potential wells provided by the density peaks of the cosmic web: the DM halos \citep{White:1978}, where it starts forming stars. In the linear regime of perturbation growth, the evolution of structure formation can be described analytically. Once the perturbations grow into the non-linear phase, there are various approximations that can be used \citep[][]{ZelDovich:1970,Gunn:1972,Gurbatov:2012}.  
A powerful alternative to these approximations is to trace structure formation either through N-body simulations, which cover only DM and further apply  semi-analytic galaxy formation models \citep[e.g.,][]{White:1991,Dalcanton:1997,Mo:1998}, or through hydrodynamic N-body simulations, which follow the coupled evolution of DM and baryons \citep[e.g.,][and the many  works that have followed]{Katz:1991,Katz:1992,Katz:1992b,Cen:1992,Navarro:1994}. 

Early studies of large box N-body simulations have revealed that cold DM halos are self-similar \citep{Navarro:1997}. Therefore, the density distributions of spherical DM halos can be completely described by only two parameters, namely: total mass, $M_{\rm h}$, and concentration, $c$. These two parameters are not independent. The correlation between them depends on the cosmological model assumed \citep[e.g.,][]{Maccio:2008} and this correlation also varies with redshift $z$ \citep[e.g.,][]{Dutton:2014}. The DM halo mass is an important predictor for fundamental galaxy properties such as total stellar mass and the star formation rate or star formation efficiency as probed by both analytic models \citep[e.g.,][]{Peacock:2000,vandenBosch:2003,Vale:2004,Conroy:2007,Conroy:2009,Guo:2010,Moster:2010,Moster:2018,Behroozi:2018} as well as cosmological hydrodynamical N-body simulations \citep[e.g.,][]{Schaye:2015,Wang:2015}. In observations, the masses and concentrations of DM halos can be constrained via models that \citet{Syer:1996} generically referred to as ``dynamical 
made-to-measure models'' that include methods based on: distribution functions, moments (e.g., Jeans models), orbits \citep[e.g., Schwarzschild orbit superposition method,][]{Schwarzschild:1979}, and particles \citep[introduced by][and now known as M2M]{Syer:1996}.

Because cold DM can only interact gravitationally, the other important physical property of DM halos is the angular momentum, $J_{\rm h}$. 
In theory, a halo acquires its 
angular momentum through tidal torques induced by the 
misaligments between the inertia tensor of the collapsing Lagrangian patch of the halo 
and the large scale tidal field \citep[tidal torque theory, TTT,][]{Hoyle:1951,Doroshkevich:1970,Fall:1980,White:1984}. In this scenario, most of the halo angular momentum (AM)
is acquired at high $z$, where perturbations growth linearly \citep{Catelan:1996}. 
This linear growth phase of $J_{\rm h}$ ends when the Lagrangian patch of the spherical overdensity reaches its maximum expansion. At later times, mergers are the dominant processes that can alter $J_{\rm h}$ \citep[e.g.,][]{Barnes:1987}.

A fundamental assumption of TTT is that DM and baryons are tightly coupled, at least at high $z$, such that their specific AM are roughly equal: $j_{\rm baryon}\simeq j_{\rm h}$. Another important assumption in galaxy formation models is that AM is conserved during the non-linear phase of perturbations growth. In this manner, the observed relation between galaxy sizes and their rotation velocities arises naturally by assuming the disk scale $R_{\rm d}$ to be equal to the virial radius $r_{\rm h}$ multiplied by the halo spin: $R_{\rm d}\sim\lambda r_{\rm h}$ \citep[e.g.,][]{Dalcanton:1997,Mo:1998,2008ApJ...672..776S,2009MNRAS.396.1675F,2011MNRAS.410.1660D}. 

\citet{Peebles:1971} introduced the spin $\lambda,$ which is an adimensional measure of halo angular momentum: $\lambda=J|E|^{\rm 1/2}/GM^{5/2}$, where $J$ is the total (baryons + DM) angular momentum of a halo, $E$ is the total energy, $M$ is the total mass, and $G$ is the gravitational constant. Using $\lambda$, halos of very different mass can be meaningfully compared. 
Early cold dark matter simulations resulted in median spin values $\overline{\lambda}\simeq0.04 - 0.05$ \citep{Barnes:1987,Efstathiou:1988,Cole:1996}. These simulations explored various cosmologies 
and found $\overline{\lambda}$ to be only mildly impacted by variations in the spectral index of the primordial power spectrum. Also, they 
found very weak trends for more massive halos to have lower $\lambda$.

Using more recent $\rm\Lambda$CDM cosmological simulations,
\citet{Bullock:2001} showed that the probability distribution function of DM halo spins can be 
well fitted by a log-normal, with only slightly lower median $\lambda$ than the earlier results: $\overline{\lambda}=0.042\pm0.006$. These authors also introduced a different definition of spin: $\lambda=J/M/(\sqrt{2}r_{\rm vir}v_{\rm vir})$, where $r_{\rm vir}$ and $v_{\rm vir}$ are the virial radius and velocity, respectively, for which they found $\overline{\lambda}=0.035\pm0.005$ and a log width $\sigma_{\rm ln\lambda}=0.50\pm0.03$. This definition of spin has become the prefered one because it does not require an estimate of the total energy. Over the years, simulations have increased in size and resolution, but the $\lambda$ distribution remained stable, at least in terms of $\overline{\lambda}$; for instance, \citet{Jiang:2019} obtained $\overline{\lambda}=0.037$ and $\sigma_{\rm ln\lambda}=0.215$. Furthermore, \citet{Benson:2017} showed that insufficient resolution leads to spurious high $\lambda$ values, which artificially increase $\sigma_{\rm ln\lambda}$. Therefore, the lower width of the distribution in more recent estimates can be attributed to significantly higher resolutions in these simulations ($\sim$10$^{\rm 6}$ particles per halo).

As expected from the anisotropic TTT \citep{Codis:2015}, the DM halo spin directions correlate with the large scale environment of galaxies \citep[e.g.,][]{Codis2012,Lopez:2021,Veena:2021}. Nevertheless, the alignment signal between the DM halo spin and the angular momentum of the galaxy disk is weak \citep[e.g.,][]{2005ApJ...627L..17B,2010MNRAS.404.1137B,2015MNRAS.452.4094D,2016MNRAS.460.3772S}.
Also, the modulus of halo spin $\lambda$ correlates very weakly or not at all with observable galaxy properties (or properties that can be derived from observations) such as (disk) size, stellar mass, or stellar angular momentum \citep[e.g.,][]{Teklu:2015,RodriguezGomez:2017,Desmond:2017,Jiang:2019,Behroozi:2021}. 
Hence, in semi-analytic models of galaxy formation \citep[e.g.,][]{2008ApJ...672..776S,2011MNRAS.410.1660D}, the spin is assumed to be uncorrelated with any other galaxy property, its value for individual galaxies being simply drawn from the universal log-normal distribution. This lack of correlations with galaxy observables implies that $\lambda$, one of the 
fundamental properties of DM halos, cannot be inferred for individual observed galaxies.

In this study, we revisit the subsample of galaxies from the Numerical Investigation of a Hundred Astrophysical Objects \citep[NIHAO,][]{Wang:2015}, analyzed in \citet{Obreja:2019}, to look for correlations between the DM halo AM and the individual galactic stellar dynamical structures. \citet{Obreja:2018,Obreja:2019} (hereafter Paper I and Paper II)  
showed that Gaussian mixture models applied to a
stellar dynamical space of normalized AM and normalized binding energy are capable of disentangling 
a multitude of dynamical structures such as thin and thick disks, classical and pseudo bulges, and stellar halos. 
Therefore, a possible interesting direction of inquiry is to look for correlations between the build-up of the dark matter 
halos and the formation of robustly defined individual galaxy components.   
We find tight correlations between $J_{\rm h}$ and the stellar dynamical disks, and between $J_{\rm h}$
and the dynamical stellar halos. These relations allow us to infer the DM halo spin from 
galaxy properties that can be derived from observations, and we carried out such an exercise for the 
Milky Way (MW). We used a combination of MW observational data 
from Gaia \citep{Gaia:2016}, re-analyzed by \citet{Hogg:2019}, and 
Apache Point Observatory Galactic Evolution Experiment \citep[APOGEE, ][]{Majewski:2017}, 
together with the Galaxy mass model of \citet{Cautun:2020} and the 
circular velocity curve of \citet{Eilers:2019} to compute the AM for the MW thin and thick stellar disk. 
Finally, by employing the relations found in simulations, we estimated the values of 
$J_{\rm h}$ and $\lambda$ for our Galaxy. 

\section{Simulated galaxy sample}

NIHAO \citep{Wang:2015} is a collection of zoom-in hydrodynamical simulations run with the N-body smoothed particle hydrodynamics (SPH) code \texttt{Gasoline2} \citep{Wadsley:2017} using a Planck cosmology \citep{Planck:2014}. This simulation sample covers five orders of magnitude in stellar mass, 
from dwarfs to MW/M31 analogs with  $\rm\sim$10$^{\rm 6}$ particles/halo 
in order to resolve the half mass radius of the galaxies.
Star formation is implemented stochastically in dense (n $>$ 10.3 cm$^{\rm -3}$) and cold 
(T $>$ 15000 K) gas, with an efficiency of 10\%,  such that a Kennicutt-Schmidt type relation is recovered. Two types of stellar feedback are included: SNe II blast-waves \citep{Stinson:2006} 
and photoheating of gas by massive stars before their SN stage \citep{Stinson:2013a}. The free parameters of the stellar feedback have been chosen such that one MW mass galaxy follows the abundance-matching stellar-to-halo mass relation \citep{Moster:2013,Behroozi:2013,Kravtsov:2018} throughout its evolution to $z=0$ \citep{Stinson:2013a}. With this fixed choice of feedback parameters, all NIHAO galaxies are within or very close to the permitted regions of abundance matching relations, 
both at low and at high $z$s \citep{Wang:2015}.
The simulations have been run assuming the initial mass function of \citet{Chabrier:2003}, along with heavy elements enrichment from  SNe Ia \citep{Thielemann:1986} and SNe II yields \citep{Woosley:1995}. 
Metals are allowed to diffuse as described in \citet{Wadsley:2008}. Gas is assumed to be in photoionization equilibrium with the $z$-dependent ultra violet background of \citet{Haardt:2012}, and to cool through metal lines and Compton scattering following \citet{Shen:2010}.

The DM halos are defined with the \texttt{Amiga Halo Finder} code \citep{Gill:2004,Knollmann:2009}, such that the virial radius $r_{\rm 200}$ encloses a sphere with average density 200 times the cosmic critical matter density. 

The NIHAO simulations have been tested extensively against observations, showing good agreement 
in terms of, for instance, the HI velocity function \citep{Maccio:2016,2019MNRAS.482.5606D}, Tully--Fisher relation \citep{2016MNRAS.457L..74D}, properties of galactic satellites and field dwarf galaxies \citep{Buck:2018a}, stellar components properties \citep{Obreja:2016,Obreja:2018,Obreja:2019,2020MNRAS.491.3461B}, 
 properties of ultra diffuse dwarfs \citep{2017MNRAS.466L...1D}. 
 
\citet{2020MNRAS.491.3461B} analyzed MW-like galaxies in the simulated sample, run with both the fiducial NIHAO and increased resolutions. In the high-resolution sample, they find values for the disk scale heights between 0.2 and 0.4 kpc for the thin disks, and 1.0 and 1.4 kpc for the thick ones, as well as global scale lengths from S\'{e}rsic + exponential fits of the stellar surface mass density between 3.9 and 5.7 kpc. Stellar masses are only slightly lower in the higher-resolution runs, while scale heights are lower by an average of 25\%. In the Tully--Fisher relation (Figure 8 of \citealt{2020MNRAS.491.3461B}), NIHAO galaxies occupy the upper envelope, probably 
showing that they are a little too compact when compared with the observations. 
Nevertheless, when compared against MW \citep[e.g.,][]{Bland-Hawthorn:2016}, they are in good agreement within their resolution limits. 
The rotation curves of the MW-like galaxies in the original NIHAO are shown in Fig.~\ref{vc_profiles} of Appendix~\ref{MWs_prop}, where the halo and stellar masses, as well as the spin parameters are also given. One simulated galaxy is particularly close to the MW's circular rotation curve $V_{\rm c}(R),$ as estimated by, for instance, \citet{Eilers:2019}, with the other MW-like galaxies wobbling around the observationally constrained $V_{\rm c}(R)$. 

The stellar metallicities of NIHAO galaxies are in agreement with the observations of \citet{Gallazzi:2005} and \citet{Panter:2008} for stellar masses larger than 10$^{\rm9}$M$_{\rm\odot}$ \citep[see][]{2021arXiv210303884B}. 
Discrepancies are mostly restricted to the dwarf regime ($M_{\rm *}<$10$^{\rm9}$M$_{\rm\odot}$), where both stellar and gas metallicities at fixed stellar mass tend to be lower in NIHAO than in the observations \citep{McConnachie:2012,Berg:2012,Kirby:2013,Zahid:2013}. Considering the limited statistics of both simulations and observations, the large overlap between the above mentioned observations and the NIHAO galaxies indicate that these simulations have realistic metallicities at fixed stellar mass\footnote{Metallicity uncertainties for observed galaxies can be quite large, for instance, 0.7 dex differences depending on the calibration \citep{Kewley:2008}. Also, the vast majority of simulation works, including the ones cited here, do not measure metallicity in the same way as in observations because it would require costly simulation post-processing with radiation transfer codes. Thus, comparisons between simulations and observations should be taken with a grain of salt.}.

The NIHAO simulations contributed to firmly establish that baryons can break the self similarity  of DM halos, which develop cores in the halo mass ranges where star formation is most efficient \citep{2016MNRAS.456.3542T,2016MNRAS.461.2658D,2017MNRAS.467.4937D}. 
This core formation effect 
predicted by \citet{1986ApJ...301...27B} 
is seen in simulations that can resolve sufficiently high gas densities \citep[e.g.,][]{2015MNRAS.454.2981C,2019MNRAS.486..655D,2020MNRAS.499.2648D}, with a strength depending on the stellar/halo mass ratio  \citep[e.g.,][]{DiCintio:2014,2016MNRAS.456.3542T}, and on the density threshold for star formation \citep[e.g.,][]{Governato:2012,2019MNRAS.488.2387B}.   
Recently, the NIHAO sample has also been extended to massive elliptical galaxies after the implementation of super massive black hole formation, accretion, and feedback in the version of the \texttt{Gasoline2} code used to run the original sample \citep{2019MNRAS.487.5476B}; the addition of AGN feedback resolved the overcooling of the three most massive NIHAO halos.   

In the current study, we used the subsample of 25 NIHAO galaxies from \citet{Wang:2015}, 
analyzed in Paper II. These galaxies have stellar masses between 7$\times$10$^{\rm 8}$M$_{\rm\odot}$ and 2$\times$10$^{\rm 11}$M$_{\rm\odot}$, and have been chosen
to show no obvious signs of interactions in the stellar surface mass density maps at $z=0$. The following section gives a brief overview of Papers I and II, focusing on our definition of stellar galactic structures.


\section{Defining the dynamical stellar structures}
\label{sec_def}

Historically, galaxies have been classified based on their disky or spheroidal photometric appearance, 
leading to the so-called Hubble diagram \citep{Sandage:1961}. With the recent advances in Integral Field Unit (IFU)
spectrographs such as PPak at the Calar Alto Telescope \citep{Kelz:2006}, Multi Unit Spectroscopic Explorer (MUSE) at the Very Large Telescope (VLT) \citep{Bacon:2010}, 
Sydney-AAO Multi-object Integral-field spectrograph (SAMI) at the Anglo-Australian Telescope \citep{Croom:2012}, or Mapping Nearby Galaxies at Apache Point Observatory (MaNGA) at Sloan \citep{Bundy:2015}, the galaxy formation community has begun to broaden and refine the classification schemes, incorporating dynamical galaxy properties that can be extracted from the modeling of IFU data.

Unlike the observational data, simulations provide direct access to the full 6D position-velocity space of stellar particles through time. Therefore, in simulations, it is relatively easier to derive robust measures of intrinsic dynamical properties, such as the ratio of ordered to random motions (or rotational support $v/\sigma$). One particularly straightforward and widely used method for looking at the extent to which a simulated galaxy is rotationally supported has been introduced by \citet{Abadi:2003}. These authors 
constructed the histograms of the stellar circularity parameter $\epsilon$ defined as the (specific) azimuthal AM $j_{\rm z}$ normalized to the (specific) AM of a circular orbit with the same binding energy $j_{\rm c}$: $\epsilon=j_{\rm z}/j_{\rm c}$, where the $z$-axis is given by the direction of the stellar AM within the galaxy region ($\lesssim0.2r_{\rm vir}$). This parameter is $-1\eqslantless\epsilon\eqslantless1$, with counter-rotating particles having $\epsilon<0$, and disk particles clustering strongly toward $\epsilon\eqslantless1$. Therefore, assuming the spheroid or bulge to have no net rotation (corresponding to a symmetric $\epsilon$ distribution around $0$), the dynamical bulge-to-total ratio can be computed as $B/T_{\rm dyn} = 2\Sigma_{\rm i} M_{\rm i}(\epsilon_{\rm i}<0)/\Sigma_{\rm i} M_{\rm i}$, where $M_{\rm i}$ is the mass of particle, $i$. Thus, a good measure of rotational to velocity dispersion support becomes the corresponding dynamical disk-to-bulge ratio.  

In order to get a less ambiguous assignment of particles as belonging to a disk or a spheroid or bulge, 
\citet{Domenech:2012} employed a modified version of the $kmeans$ cluster-finding algorithm in a 3D stellar 
kinematic space of circularity, normalized in(equatorial)-plane AM, $j_{\rm p}/j_{\rm c}$ (with total specific AM: 
$\vec{j}=\vec{j_z}+\vec{j_p}$), and binding energy, $E$. In this way, they were able to separate not only the disk from the 
bulge, but also to pick up two different disk types: a thin and a thick one. 

In Paper I, we refined the method introduced by \citet{Domenech:2012}, switching to Gaussian mixture models
with unconstrained covariance instead of $kmeans$, in order to relax the latter algorithm's limitation to convex clusters of 
roughly equal weights. The 3D feature space that we use is ($j_z/j_c$,$j_p/j_c$, $e/|e|_{\rm max}$), where the specific
binding energy $e$ is normalized to the most bound stellar particle in the halo $|e|_{\rm max}$. This method is capable of disentangling a multitude of dynamical structures in simulated galaxies, including components that resemble observed thin or thick disks, classical or pseudo-bulges, and stellar halos (Paper II). The code we developed to select (in this manner) the dynamical components of simulated galaxies is called \texttt{galactic structure finder (gsf)}\footnote{available at \url{https://github.com/aobr/gsf}. The new version of the code, \texttt{gsf2}, is available upon request from the main author. This version includes the possibility to choose other spaces for clustering, provides statistics for a wide range of physical parameters of each component, and uses an information criteria to automatically select the optimal number of components in a galaxy.}. To get meaningful results, we tailored \texttt{gsf} to work on relaxed galaxies and to recompute the gravitational force at each stellar particle position in a galaxy by assuming its halo to be in isolation.
In this context, ``relaxed'' means that the symmetry $z$-axis can be defined unambiguosly; the direction of the total stellar angular momentum computed within spheres of increasing radii has to vary little if at all with the radius.
In this way, the normalized specific binding energy $e/|e|_{\rm max}$ is ensured to be well behaved for all stellar particles in a halo. 

The relatively high resolution of the NIHAO simulations allowed us to go after dynamical galaxy components as faint as what is typically expected from stellar halos. To choose the optimal number of components, one straightforward strategy is to construct the functions Bayes Information Criteria \citep[BIC,][]{Schwarz:1978} or Akaike Information Criteria \citep[AIC,][]{Akaike:1974} versus the number of Gaussians. In the \texttt{scikit-learn} Python library \citep{Pedregosa:2011} that \texttt{gsf} uses for the clustering, the BIC is defined as: $BIC=-2N\cdot ln(L)+n_{\rm param}\cdot ln(N)$, and the $AIC$ is $AIC=-2N\cdot ln(L)+2n_{\rm param}$, where $N$ is the total number of samples (particles in this case), $ln(L)$ is the per-sample average log likelihood, and $n_{\rm param}$ is the number of free parameters, directly proportional to the number of Gaussian components $n_{\rm k}$: $n_{\rm param}=10n_{\rm k}-1$ for a 3D feature space 
and fully unconstrained covariance matrix. The large number of samples (stellar particles) $N$ to which we apply the Gaussian mixture models makes both of these criteria just scaled version of the log likelihood, since the penalty functions $n_{\rm param}\cdot ln(N)$ and $2n_{\rm param}$ are much smaller that the first $2N\cdot ln(L)$ terms. A typical curve $-ln(L)$ vs $n_{\rm k}$ for a galaxy analyzed with \texttt{gsf} is shown in \citet{Buck:2019}, and has no clear minimum that would permit selecting the "optimal"\ number of 3D Gaussians in a straightforward manner (however, also see \citealt{Du:2019}, who used \texttt{gsf} with a heuristic BIC criteria to select the number of components in simulated galaxies).

The goal in Paper II was to identify stellar halos and separate disks if possible into thin and thick components. Therefore, visual inspection of models with similar $ln(L)$/$BIC$ was necessary, given the large overlap in feature space of the 
various galaxy components and the small weight of stellar halos. The smaller (dwarf) galaxies in the sample have been separated into only two components: a disk and a spheroid, while the MW analogs in terms of stellar mass are made of up to five different dynamical structures (see their Figure A1). Table~\ref{table:1} gives the positions of the following distinct dynamical components in the \texttt{gsf} input feature space: 
spheroid, classical bulge, pseudo bulge, (stellar) halo, (single large-scale) disk, thin disk, and thick disk.
Spheroids are very close to spherical symmetry, can have large radial extents, are very bound, have circularities peaking close to 0, and show no traces of rotation in the line-of-sight edge-on perspective. Classical bulges are very close to spherical symmetry, very bound, very compact, have circularities peaking close to 0, and show no traces of rotation in the line-of-sight edge-on perspective. Pseudo-bulges are flattened, compact, less bound than classical bulges and spheroids, and show signals of rotation in the line-of-sight edge-on perspective. Stellar halos are the least bound components, with very large radial extents, and usually show some signs of rotation (their peak circularities are slightly higher than 0). Single large-scale disks are flat, extended, with strong rotation patterns in the line-of-sight edge-on perspective, and their circularities peak at large values. Thin disks are very flat and extended, show  strong rotation patterns in the line-of-sight edge-on perspective, and have the largest circularity peak -- close to 1 -- among the components of galaxies. Thick disk are less flat and extended than thin disks, show  significant rotation patterns in the line-of-sight edge-on perspective, and their circularities peak at values lower than their corresponding thin disks.
The corresponding values for all the components of all the galaxies in the sample are given in the table A1 of Paper II. The components' names have been chosen based on their surface mass density and line-of-sight velocity maps, as explained above, and on their positioning in the ($j_z/j_c$,$j_p/j_c$, $e/|e|_{\rm max}$) space.

\begin{table}
\caption{Positions of the dynamical components in the feature space}
\label{table:1}
\centering
\begin{tabular}{c c c c}
\hline\hline
Component & $j_{\rm z}/j_{\rm c}$ & $j_{\rm p}/j_{\rm c}$ & $e/|e|_{\rm max}$\\
\hline
spheroid & 0.05 $\rm\pm$ 0.13 & 0.41 $\rm\pm$ 0.07 &  -0.76 $\rm\pm$ 0.03\\

classical bulge &  0.02 $\rm\pm$ 0.08 &  0.48 $\rm\pm$ 0.04 &  -0.78 $\rm\pm$ 0.03\\

pseudo bulge &  0.18 $\rm\pm$ 0.09 &  0.17 $\rm\pm$ 0.02 &  -0.71 $\rm\pm$ 0.03\\

halo &  0.07 $\rm\pm$ 0.09  & 0.35 $\rm\pm$ 0.06 &  -0.39 $\rm\pm$ 0.04\\

disk &  0.68 $\rm\pm$ 0.08 &  0.24 $\rm\pm$ 0.04  & -0.61 $\rm\pm$ 0.04\\

thin disk  & 0.88 $\rm\pm$ 0.02 &  0.18 $\rm\pm$ 0.03 &  -0.61 $\rm\pm$ 0.03\\

thick disk  & 0.64 $\rm\pm$ 0.05 &  0.33 $\rm\pm$ 0.04 &  -0.68 $\rm\pm$ 0.03\\

\hline
\end{tabular}
\end{table}

As already discussed in Paper II and shown in Table~\ref{table:1}, the stellar dynamical components are well ordered in binding energies, from stellar halos, which are the least bound, passing through the disks, thin disks, thick disks, pseudo bulges, spheroids, and, finally, to classical bulges, which are the most bound structures. In terms of circularities, the most rotational supported components are the thin disks, followed by the disks, thick disks, and pseudo bulges, while the stellar halos, spheroids and classical bulges have almost no net rotation. 

\begin{figure*}
\centering
\includegraphics[width=1.0\textwidth]{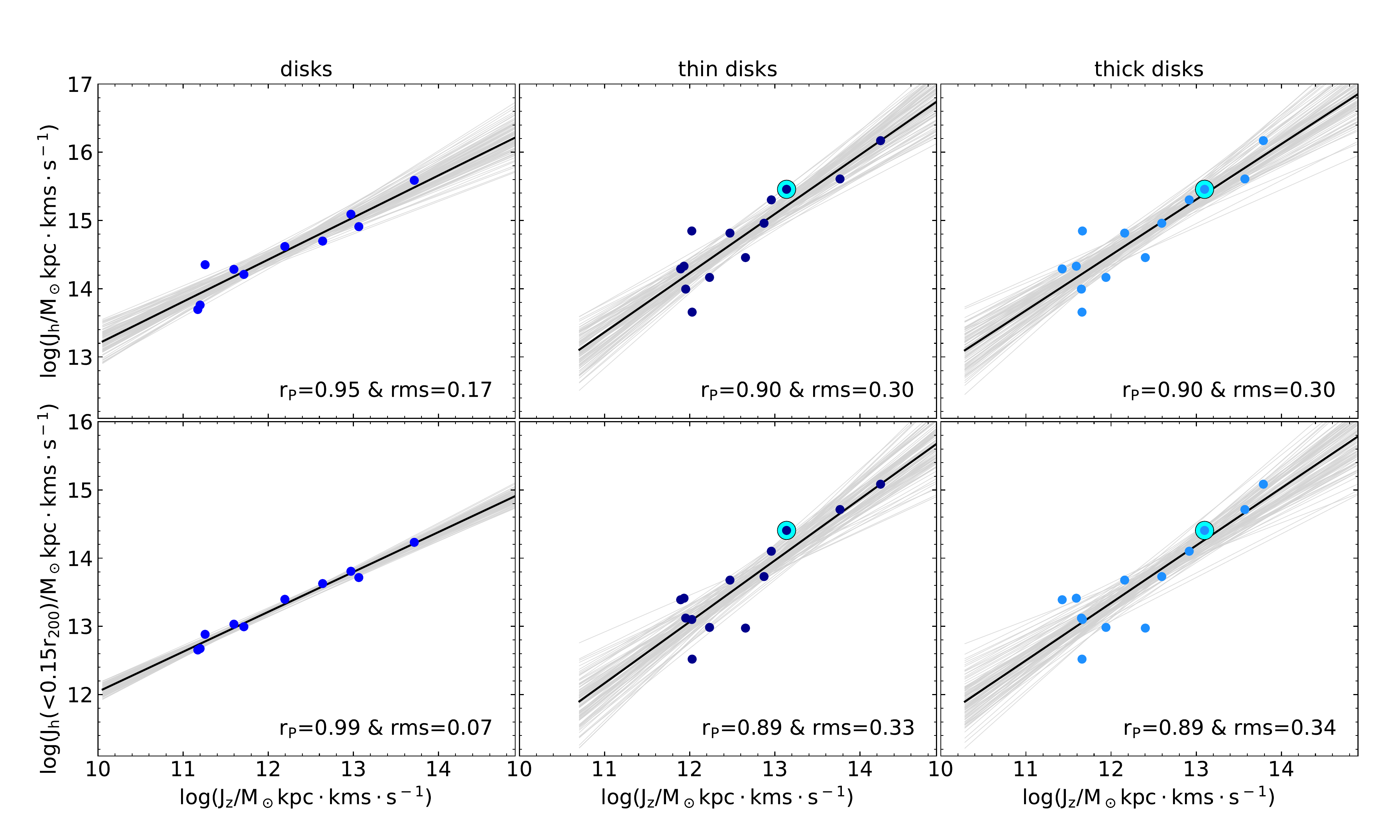}
\caption{Relations between the azimuthal AM of the dynamical disk components and the AM modulus of the DM halo for the galaxy sample in Paper II. In the top panels the y-axis gives the total DM AM within the virial radius $J_{\rm h}$, while the bottom ones show the AM within 15\% of the virial radius $J_{\rm h}(<0.15r_{\rm 200})$. From left to right, the panels show the relations for (single large scale) disks, thin and thick, respectively. In each panel, the black line is the linear fit through the data points, while the gray lines show 100 realization of the relation drawn from the covariance matrix of the fit. The Pearson correlation coefficient $r_{\rm P}$ and the 
root-mean-square deviation $rms$
are given in the bottom right corner of each panel. In the top and central panels, the highlighted cyan points correspond to the best MW analog in the NIHAO sample, g8.26e11 \citep{Obreja:2018,2020MNRAS.491.3461B}.} 
\label{Jdark_vs_Jdisks}
\end{figure*}

\begin{figure*}
\centering
\includegraphics[width=0.67\textwidth]{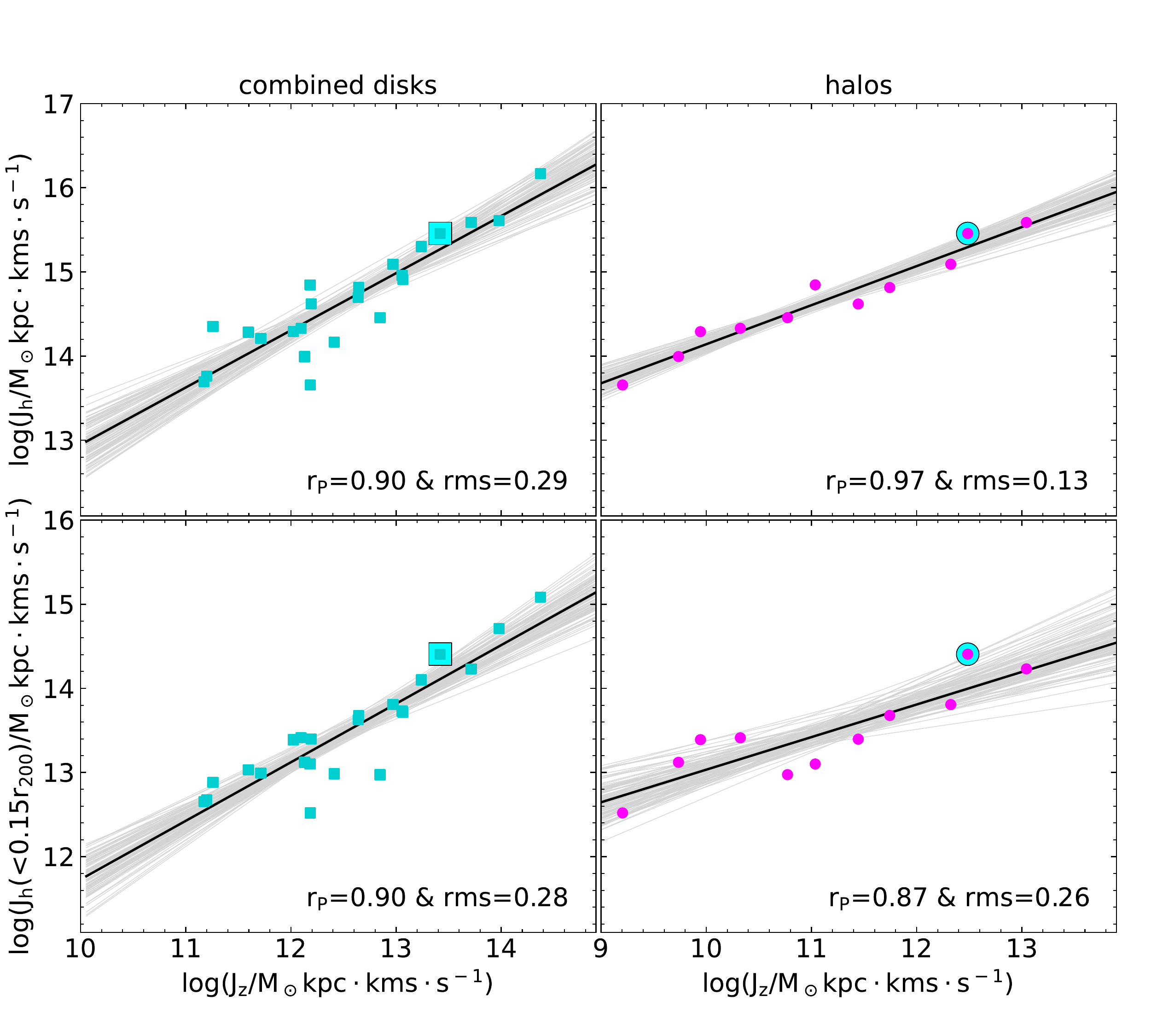}
\caption{Relations between the azimuthal AM of the dynamical disk components and the AM modulus of the DM halo for the galaxy sample in Paper II. Same details as in Fig.~\ref{Jdark_vs_Jdisks}, but for all the dynamical disks combined (left) and the stellar halos (right). In the left panels, each galaxy with a double disk is represented 
by a single point with abscissa $J_{\rm z}=J_{\rm z,thin}+J_{\rm z,thick}$. The highlighted cyan points in all panels correspond to the best MW analog in the NIHAO sample, g8.26e11.} 
\label{Jdark_vs_Jdisksandhalos}
\end{figure*}


\section{$J_{\rm h}$ vs the AM of stellar disks and halos}
\label{sec_relations}

In Paper II, we discussed at length the evolution of the specific AM for the various types of 
stellar structures present in the simulations. In particular, by tracing back in time the 
Lagrangian masses of the $z=0$ stellar structures and DM halos, we were able to assess the differences between the AM evolution of galactic components and that of their DM halos. 
Thus, it resulted that stellar disks (thin or thick, or single large-scale disks) have the most similar 
evolution to the DM halos (figure 12 in Paper II). On average, the material of $z=0$ disks reached 
its peak AM at redshifts $z_{\rm peak}\sim1.37$ and $\sim1.65$ for thin and thick disks, respectively, and $z_{\rm peak}\sim2.11$ for single large-scale disks, while the material of the host halos peaked in AM at $z_{\rm peak}\sim1.44$. From $z_{\rm peak}$ to $z=0$, disks and DM halos lose some part of the maximum AM reached, with the DM halos retaining the most $f_{\rm j}=0.75\pm0.15$, and the thick disks the least $f_{\rm j}=0.41\pm0.15$, where $f_{\rm j}$ quantifies the retained fraction of angular momentum as the ratio between the (specific) AM at $z=0$ and the maximum specific (AM) reached during the universe's lifetime: $f_{\rm j}=j(z=0)/max(j)$. By comparison, the dispersion-dominated structures (classical and pseudo bulges, and spheroids) have larger $z_{\rm peak}$ and they lose most of their acquired AM, having an overall AM evolution that is not particularly similar to the DM one. Also, it is important to take into account that supernova feedback can remove low AM gas from galaxies \citep[e.g.,][]{Dutton:2009a,Brook:2011}, some of which will come back onto the symmetry plane at larger radii and with larger AM acquired in the likely spinning circumgalactic medium \citep[e.g.,][]{2006MNRAS.366..449F,Brook:2012,Athanassoula:2016,Peschken:2017}. This recycling of disk material prior to the star formation which happens preferentially in or very close to the symmetry plane, together with secular processes such as radial migration \citep[][]{2002MNRAS.336..785S}, means that it is not expected for stellar disks and dark matter to precisely share  the same specific AM.

The similarities in AM evolution between the progenitor material of stellar disks and that of DM halos do indeed translate into 
relatively tight correlations at $z=0$ between the 
azimuthal AM component of stellar disks and the AM modulus of the DM halo, 
as shown in  Fig.~\ref{Jdark_vs_Jdisks}. The top panels
give the total DM halo AM measured within the virial radius $r_{\rm 200}$, while the bottom ones the corresponding DM halo AM measured in the galaxy region $r<0.15r_{\rm 200}$. We fitted all these individual relations with power laws: 
\begin{equation}
 log(J_{\rm h}) = \alpha + \beta\cdot log(J_{\rm z})
 \label{eq_J}
,\end{equation}
which are depicted by the solid black lines in the panels of Fig.~\ref{Jdark_vs_Jdisks},
where both $J_{\rm h}$ and $J_{\rm z}$ are measured in M$_{\rm\odot}$kpc$\cdot$km$\cdot$s$^{\rm -1}$. The Pearson correlation coefficient $r_{\rm P}$ as well as the root-mean-square deviation $rms$
are quoted in each panel, and all the fit parameters are given in Table~\ref{table:2}. We chose to plot on the x-axis the azimuthal component of the stellar AM, given that this is the quantity that can be estimated from observations (for disks $J \cong J_{\rm z}$), while $J_{\rm h}$ plotted on the y-axis is the quantity we want to estimate.

\begin{table*}
\caption{Best-fit parameters for the relations between the AM of the DM halo and that of the stellar dynamical components.}
\label{table:2}
\centering
\begin{tabular}{c | c c c c | c c c c }
\hline\hline
& \multicolumn{4}{|c|}{log($J_{\rm h}$)=$\alpha$+$\beta\cdot$log($J_{\rm z}$)} & \multicolumn{4}{|c|}{log($J_{\rm h}$($<0.15r_{\rm 200}$))=$\alpha$+$\beta\cdot$log($J_{\rm z}$)}\\
\hline
Component & $\alpha$ & $\beta$ & $r_{\rm P}$ & $rms$ & $\alpha$ & $\beta$ & $r_{\rm P}$ & $rms$\\
\hline
\rowcolor{lightgray} halos & 9.50$\rm\pm$0.42 & 0.46$\rm\pm$0.04 & 0.97 & 0.13 & 9.16$\rm\pm$0.82 & 0.39$\rm\pm$0.07 & 0.87 & 0.26\\
disks & 7.03$\rm\pm$0.88 & 0.62$\rm\pm$0.07 & 0.95 & 0.17 & 6.19$\rm\pm$0.35 & 0.58$\rm\pm$0.03 & 0.99 & 0.07\\
thin disks & 3.83$\rm\pm$1.58 & 0.87$\rm\pm$0.13 & 0.90 & 0.30 & 2.27$\rm\pm$1.75 & 0.90$\rm\pm$0.14 & 0.89 & 0.33\\
thick disks & 4.74$\rm\pm$1.47 & 0.81$\rm\pm$0.12 & 0.90 & 0.30 & 3.24$\rm\pm$1.64 & 0.84$\rm\pm$0.13 & 0.89 & 0.34\\
\rowcolor{lightgray} combined disks & 6.15$\rm\pm$0.92 & 0.68$\rm\pm$0.07 & 0.90 & 0.29 & 4.77$\rm\pm$0.91 & 0.70$\rm\pm$0.07 & 0.90 & 0.28\\
\hline
\end{tabular}
\end{table*}

When looking at the total DM halo AM within the virial radius $J_{\rm h}$, the 
tightest correlation appears for the single disks ($r_{\rm P}=0.95$ and $rms=0.17$). 
We attribute this behavior to the nearly perfect coevolution of angular momenta of single disks' progenitor material and DM halos' progenitor material up to $z\backsimeq2$, as shown by Fig. 12 in Paper II (blue versus thick black). 
The thin and thick disks have a slightly different high-$z$ AM history than the DM halos, translating into a less tight correlation and larger scatter, even if their low-$z$ evolution is more similar to that of DM halos than to that of single disks. If the AM of the DM halos is measured within the galaxy region 
($r<0.15r_{\rm 200}$), the correlation with the single disks is very close to ideal ($r_{\rm P}=0.99$ bottom-left panel of Fig.~\ref{Jdark_vs_Jdisks}), while for the thin and thick disk, it is slightly worse (central and right bottom panels versus central and right top ones). These differences in behavior between single and  double stellar disk galaxies is worth looking into more detail in a future study, since they possibly encode information on why some objects develop one instead of two disk components.

In observations, it is not always straightforward or possible to classify a galaxy as having either a thin + thick disk or a single disk \citep[e.g.,][]{2017ApJ...847...14E}. This fact set together with our small galaxy sample size  naturally leads to the left panels of Fig.~\ref{Jdark_vs_Jdisksandhalos}, where we considered  the single and double (thin + thick) disks together in order to increase the statistics. The parameters of the power law fits shown in this figure are also given in Table~\ref{table:2}. The relation between the AM of stellar disks and that of their DM halos has a direct
application in observations; many works have already quantified the AM of large samples of stellar disks in order to look at how the AM scales with the stellar mass \citep[e.g.,][]{Fall:1980,Fall:2018,Romanowsky:2012,2014ApJ...784...26O,2018A&A...612L...6P}. In this manner, spin 
distribution for observed galaxy samples could be computed and compared with the predictions 
from simulations. For instance, \citet{Teklu:2015} found that halos hosting disk-dominated galaxies have the log-normal distribution of DM halo spins  skewed toward larger $\lambda$ than the full DM halo population (0.064 vs 0.043). In the next section, we apply the relation for combined disks 
to MW to estimate the spin of its host halo, and leave for a future work deriving the $\lambda$ distribution for large samples of observed galaxies.

In the right panel of Fig.~\ref{Jdark_vs_Jdisksandhalos}, we show the only tight correlation we found within the group of dispersion dominated stellar components, namely between the stellar halos azimuthal AM and their DM halos. Actually, at the virial scale, the DM halo AM can be most accurately predicted from the $J_{\rm z}$ of stellar halos ($r_{\rm P}=0.97$ and $rms=0.13$). The main reason for this correlation is the comparable spatial extent of stellar and DM halos. 
As can be appreciated from Fig.~\ref{Jdark_vs_Jdisksandhalos}, and from the differences in Table~\ref{table:2} (between the normalization of the relation involving the complete DM halo extent and the galaxy region only), most of the contribution to $J_{\rm h}$ comes from radii $R>0.15r_{\rm 200}$. Therefore, it is natural that the AM of the extended stellar halos would closely predict the AM of the extended DM ones. Though it is the tightest, the relation between the stellar and DM halos is harder to use in observations, because it requires: i) accurate velocity tracers all throughout a very faint galaxy component that can possibly extend all the way to the virial radius, and ii) accurate mass distributions for the dynamically distinct stellar halo.


\section{Application to MW}

The MW is now known to host a large diversity of substructure thanks to the many 
observational campaigns that targeted its stars, such as: Radial Velocity Experiment \citep[RAVE, ][]{2006AJ....132.1645S}, APOGEE \citep{2008AN....329.1018A}, Sloan Extension for Galactic Understanding and Exploration \citep[SEGUE, ][]{2009AJ....137.4377Y}, Large Sky Area Multi-Object Fibre Spectroscopic Telescope \citep[LAMOST, ][]{2012RAA....12..735D}, Gaia-ESO Survey \citep{2012Msngr.147...25G}, GIRAFFE Inner Bulge Survey \citep[GIBS][]{2014A&A...562A..66Z}, 
or Galactic Archaeology with the HERMES survey \citep[GALAH,][]{2015MNRAS.449.2604D}. 
The main body of stars of the Galaxy are structured into: a nuclear star cluster \citep{Becklin:1968}, a boxy/peanut bulge \citep{Okuda:1977}, a bar \citep{Hammersley:2000}, a thin and a thick disk \citep{Gilmore:1983},  and a stellar halo \citep{Searle:1978}. There are some indications that 
the Galaxy might also host a small classical bulge \citep[e.g., discussion in the review by][]{Bland-Hawthorn:2016}. While the MW's proximity allows the mapping of individual stars, and an unmatched level of detail with respect to other galaxies, it also obfuscates estimates of some key parameters such as the scale lengths of the thin and thick disk.  

The thick disk of the MW separates from the thin one in various properties. The thick disk has a larger scale height \citep[e.g.,][]{2008ApJ...673..864J}, older stellar ages \citep[e.g.,][]{1995AJ....109.1095G} and it is less rotationally supported \citep[e.g.,][]{2003A&A...398..141S}, more metal poor \citep[][]{Fuhrmann:1998}, and more $\alpha$ element-enhanced \citep[e.g.,][]{Fuhrmann:1998,Bensby:2003,2006MNRAS.367.1329R} than the thin disk. 

For the purpose of deriving the AM and spin of MW's DM halo , we need to first compute the AM of its thin and thick stellar disks. Under the assumption of axis symmetry, $J_{\rm z}$ can be computed in observations as:
\begin{equation}
 J_{\rm z} = \int_0^{\infty}\int_{-\infty}^{\infty}\rho(R,z)v_{\rm\phi}(R,z)2\pi R^2 dR dz,
 \label{eq_Jz}
\end{equation}
where $R$ and $z$ are cylindrical coordinates, $\rho(R,z)$ is the 3D mass density distribution, 
and $v_{\rm\phi}(R,z)$ is the rotational velocity. For galaxy disks, a usual assumption is that
$\rho(R,z)$ takes the form:
\begin{equation}
 \rho(R,z)=\frac{\Sigma_0}{2 z_d} exp\left(-\frac{R}{R_d}\right) exp\left(-\frac{|z|}{z_d}\right),
\end{equation}
with $R_d$ and $z_d$ as the radial and vertical scale lengths and $\Sigma_0$ the normalization of the 
face-on exponential surface mass density profile. Under the additional assumption that the rotational
velocity does not depend on the height above the plane $v_{\rm\phi}(R,z)=v_{\rm\phi}(R)$, Eq.~\ref{eq_Jz} becomes:
\begin{equation}
 J_{\rm z} = \frac{M}{R_d^2} \int_0^{\infty}v_{\rm\phi}(R)e^{-R/R_d} R^2 dR,
 \label{eq_Jz_approx}
\end{equation}
where $M=2\pi R_d^2\Sigma_0$ is the total disk mass. 

Therefore, to compute the AM of the MW's disk in the approximation of Eq.~\ref{eq_Jz_approx}, we need the total mass $M$, exponential scale length, $R_d$, and the rotational 
velocity profile ,$v_{\rm\phi}(R)$. For both the thin and thick disk of the MW, we can take the total mass and 
scale length from previous published works. Therefore, in order to compute $J_{\rm z,thin}$ and $J_{\rm z,thick}$, we only need to construct the velocity profiles $v_{\rm\phi,thin}(R)$ and $v_{\rm\phi,thick}(R)$.

\begin{figure*}
\centering
\includegraphics[width=1.0\textwidth]{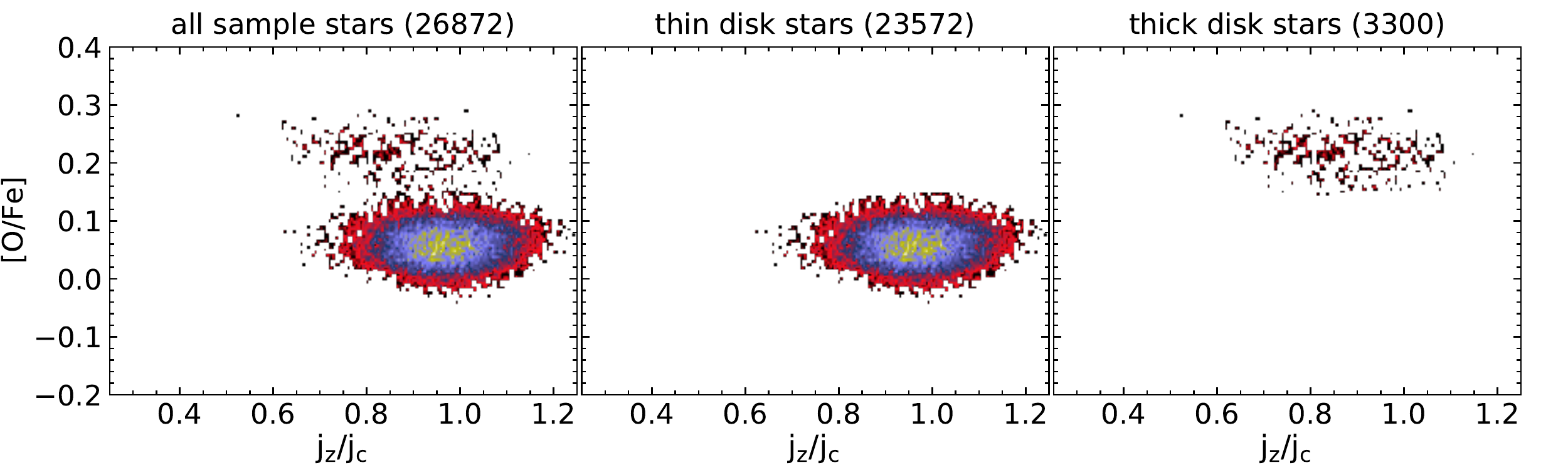}
\caption{MW disk separation in the feature space ($j_{\rm z}/j_{\rm c}$,[O/Fe]) 
using Gaussian Mixture clustering with two components. The observational data are a subsample of the stars in \citet{Hogg:2019} for which the circularity parameter $j_{\rm z}/j_{\rm c}$ could be estimated. We approximate $j_{\rm c} \cong V_{\rm c}R$, where $V_{\rm c}$ is the circular velocity at the radial position $R$ of the star in the galactic plane, and $V_{\rm c}(R)$ is taken from \citep{Eilers:2019}.}  
\label{MW_jzjcOFe}
\end{figure*}

\begin{figure*}
\centering
\includegraphics[width=1.0\textwidth]{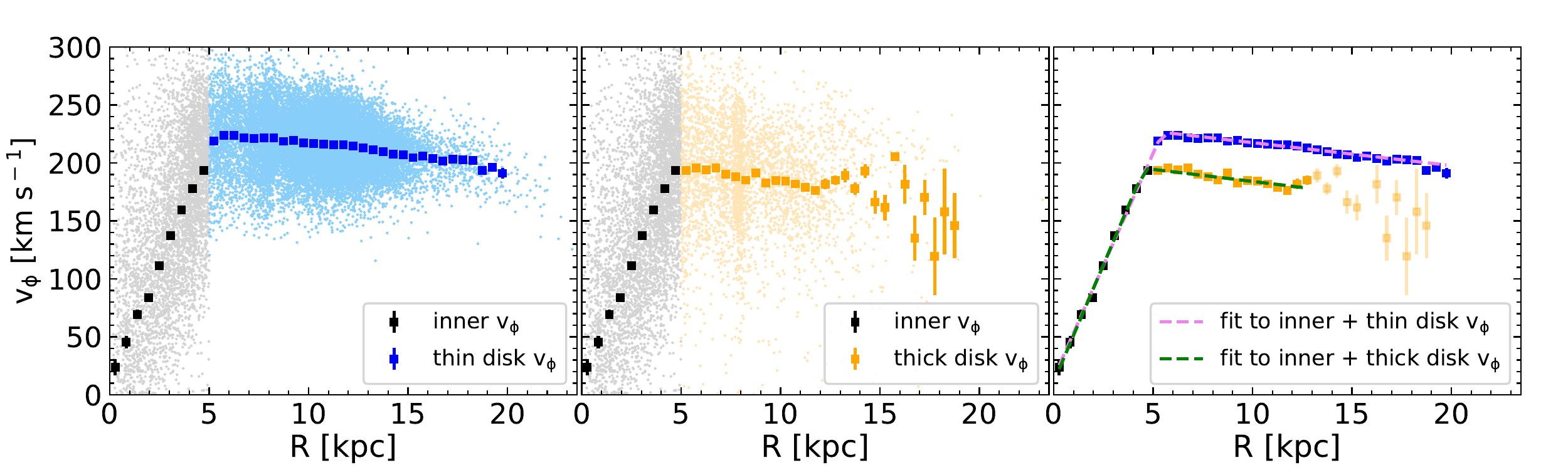}
\caption{Rotational velocity profiles for the MW's disks. The gray points in the left and central panels are the stars with $|z|<1$~kpc and $|v_{\rm\phi}|<350$~km~s$^{\rm -1}$, which were used to construct the inner $v_{\rm\phi}$ profile. The light blue and orange points in the left and central panels, respectively, are the thin- and thick-disk stars selected in the ($j_{\rm z}/j_{\rm c}$,[O/Fe]) space. The parameters of the two fits in the right panel are given in Table~\ref{tab_fit_vphi}. For the thick disk, only the region $R<13$~kpc was considered for the fit.} 
\label{MW_velprofiles}
\end{figure*}

\subsection{Thin and thick disks of the MW}
\label{vel_profiles}

To separate the MW's stellar disk into a thin and a thick component, we 
used the sample of red-giant branch stars with parallaxes re-estimated by 
\citet{Hogg:2019}. These authors used stars from a cross-matched catalog 
of APOGEE, Gaia, Two Micron All-Sky Survey \citep[2MASS, ][]{Skrutskie:2006} and Wide-field Infrared Survey Explorer \citep[WISE, ][]{Wright:2010} to construct a data-driven model capable of predicting parallaxes with less than 15\% uncertainties, therefore improving on the values available from Gaia. Their model implicitly accounts for reddening and extinction, and results in precise distances to stars with less than 10\% uncertainties up to 20 kpc from the Sun. These characteristics make this sample ideal to construct dynamical models of the Galaxy \citep[e.g.,][]{Eilers:2019,2020ApJ...900..186E,Cautun:2020}, and are very well suited for our purpose of computing the rotational velocity profiles for the thin and thick disks.

One way to separate MW's thin disk from the thick one would be to apply a clustering algorithm in the same parameter space as the one used by \texttt{gsf}. This approach, however, is complicated by the fact the Galaxy's mass models still have large uncertainties, and these models are needed to compute the stellar binding energies.  
If we restrict ourselves only to the circularity parameter, $j_{\rm z}/j_{\rm c}$, we can use circular velocity curves $V_{\rm c}(R)$ from literature to compute the normalization, $j_{\rm c}$. For this purpose, we used the 
circular velocity curve published by \citet{Eilers:2019}, who derived it by fitting 
a Jeans model to the star sample with improved parallaxes of \citet{Hogg:2019}, and radial velocities and positions from APOGEE. Therefore, we approximate
$j_{\rm c} \cong V_{\rm c}R$, where $V_{\rm c}$ is the circular velocity at 
the radial position, $R,$ of the star in the galactic plane. 
The circular velocity curve of Eilers et al. does not extend to $R<5$~kpc because their modeling cannot properly account for the presence of the bar \citep[e.g.,][]{2015MNRAS.450.4050W}. For this reason, we separate the thin from the thick disk only in the radial range $R>5$~kpc. Given the extensive evidence that the MW's disk stars show a bimodality in their $\alpha$-enhancement \citep[e.g.,][]{Fuhrmann:1998,Bensby:2003}, 
we chose [O/Fe] from the APOGEE catalog as second feature for the clustering.
We chose oxygen as a tracer of $\alpha$-enhancement for two reasons: 
i) it results in a clean separation of the two MW disk components and ii) it facilitates comparisons with our previous simulations, where only oxygen and iron are traced in detail. Using other elements (e.g., magnesium) or combinations of $\alpha$-elements available in APOGEE does not change the conclusions of this study. 

To convert the positions and velocities of stars to galactocentric coordinates
with Sgr A$^{\rm *}$ as origin, we assumed:
a position of the Sun $(x_{\rm\odot}$,$y_{\rm\odot}$,$z_{\rm\odot})=(R_{\rm\odot},0,z_{\rm\odot})$~kpc, where $R_{\rm\odot}=8.122$~kpc \citep{2018A&A...615L..15G} and $z_{\rm\odot}=0.025$~kpc \citep{2008ApJ...673..864J}, and a velocity of the Sun 
$(vx_{\rm\odot}$,$vy_{\rm\odot}$,$vz_{\rm\odot})=(-11.1,245.8,7.8)$~km$\cdot$s$^{\rm -1}$ \citep{2004ApJ...616..872R}, which is the same as in \citet{Eilers:2019}.

We made further selections to the cross-matched APOGEE and \citet{Hogg:2019} catalogs to best suit 
our purpose. Thus, we chose only: i) prograde stars with $0<v_{\rm\phi}<350$~km~s$^{\rm -1}$, ii)
confined within 1~kpc from the galactic plane $|z|<1$~kpc, and iii) with positive defined errors for [O/Fe]. 
The first two selection criteria ensure we are minimizing the contamination from the stellar halo. The circular velocity curve of \citet{Eilers:2019} does not extend to $R<5$~kpc, further reducing the sample for clustering to 27044 stars. Figure~\ref{MW_jzjcOFe} shows the separation of the selected sample of stars in the feature space ($j_{\rm z}/j_{\rm c}$,[O/Fe]) in form 
of 2D histograms. To define the two clusters visible in the complete sample on the left panel we used 
the Gaussian Mixture clustering of \texttt{scikit-learn} with two components. This clustering method is able to separate highly imbalanced clusters. The central and right panels show the 2D histograms of the two clusters identified as the thin and thick disks, respectively. Given that the stars are clearly separated in this space, other methods such as hierarchical clustering lead to a very similar assignment of stars to one group or the other. 

\begin{figure*}
\centering
\includegraphics[width=1.0\textwidth]{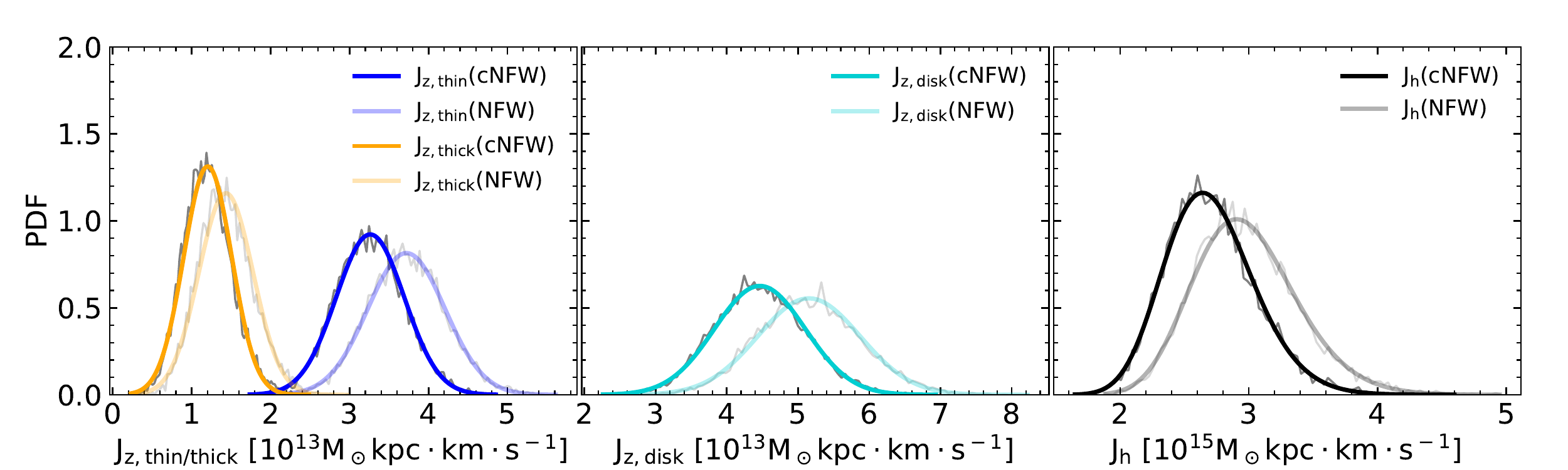}
\caption{MW angular momentum PDFs for its: thin and thick disks (left), combined stellar disk (center), and DM halo (right). The underlying PDFs in thin black are well approximated by normal distributions for the stellar components (thick blue, orange, and turquoise curves) and by log normal for the DM ones (thick black and gray curves). Solid or faint colors show the predictions for the contracted or uncontracted NFW mass model, respectively.} 
\label{AM_distributions}
\end{figure*}

Once the stars are assigned to either the thin or the thick disk, we can construct, for both components, 
the rotational velocity profiles $v_{\rm\phi}(R)$ as ensemble averaged values in fixed radial bins
of $\Delta R=0.5$~kpc for $5\leqslant R\leqslant 20$~kpc. As done in \citet{Eilers:2019}, we estimate 
the uncertainties on $v_{\rm\phi}(R)$ by bootstrapping with 100 samples. The rotational velocity 
of disk stars is expected to get to $0$ at $R=0$. Therefore, to extend the two curves below 5 kpc, 
we construct the $v_{\rm\phi}(R)$ for all stars in the cross-matched catalog with $|z|<1$~kpc and $|v_{\rm\phi}|<350$~km~s$^{\rm -1}$, and take it to represent both the thin and the thick disk rotation. 
This is a reasonable assumption considering the fact that 
exponential disks (as assumed in the mass model of \citealt{Cautun:2020}) should have $v_{\rm\phi}\rightarrow0$ as $R\rightarrow0$.
Figure~\ref{MW_velprofiles} shows the derived $v_{\rm\phi}(R)$ profiles for the thin (solid blue squares, left panel) and thick (solid orange squares, central panel) disks. Both profiles are declining with radius according to $\sim$2~km~s$^{\rm -1}$~kpc$^{\rm -1}$.
In the solar neighborhood $R=R_{\rm\odot}$, the resulting $v_{\rm\phi}$ values for the thin and thick disks of 221.2$\pm$0.8 and 188$\pm$3.4~km~s$^{\rm -1}$ are close to the values derived by \citet{Haywood:2013}:  $\rm\sim$220 and 170$\rm\pm$16~km~s$^{\rm -1}$. This means that at $R_{\rm\odot}$, the thick disk is lagging behind the thin one by $\sim$33~km~s$^{\rm -1}$.

The solid black squares in all panels of Fig.~\ref{MW_velprofiles} give the $v_{\rm\phi}(R)$ for $R<5$~kpc, computed as explained above. 
We fit the two observational profiles (solid black + solid blue representing the $v_{\rm\phi}(R)$ of the thin disk, and solid black + solid orange representing the $v_{\rm\phi}(R)$ of the thick disk) with double linear fits:
\begin{equation}
v_{\rm\phi}(R) =
\begin{cases}
v_{\rm 0}+\gamma_{\rm 0}\cdot R, & \text{ if } R < R_{\rm break}
\\
v_{\rm 1}+\gamma_{\rm 1}\cdot R, & \text{ if } R \geqslant R_{\rm break},
\end{cases}
\label{eq_vel}
\end{equation} 
where continuity at the break radius $R_{\rm break}$ implies that $v_{\rm 0} = v_{\rm 1}+(\gamma_{\rm 1}-\gamma_{\rm 0})\cdot R_{\rm break}$. The right panel of Fig.~\ref{MW_velprofiles} shows together the fits (dashed violet or green curves) and data (black and blue or orange squares) for both disks (thin or thick), and Table~\ref{tab_fit_vphi} gives the values of the parameters in Eq.~\ref{eq_vel}. The sample thick disk stars cover very sparsely regions at large radii and, therefore, the fit for the thick disk was limited to the data with $R<13$~kpc (solid vs faint orange in the right panel of Fig.~\ref{MW_velprofiles}). We did not force the fit to pass through the origin, and as a consequence $v_{\rm 0}>0$ for both disks. Setting $R_{\rm break}=5$~kpc and $v_{\rm 0}=0$ for both disks and fitting only the second branch in Eq.~\ref{eq_vel}, we get $v_{\rm 1,thin}=236.71\pm0.64$~km~s$^{\rm -1}$, $\gamma_{\rm 1,thin}=-1.93\pm0.06$~km~s$^{\rm -1}$~kpc$^{\rm -1}$, $v_{\rm 1,thick}=206.93\pm3.14$~km~s$^{\rm -1}$, and $\gamma_{\rm 1,thick}=-2.30\pm0.35$~km~s$^{\rm -1}$~kpc$^{\rm -1}$, where the continuity condition at $R_{\rm break}$ gives $\gamma_{\rm 0}=\gamma_{\rm 1}+v_{\rm 1}/R_{\rm break}$. In this case there is a more marked difference between the velocity decline with radius, $\gamma_{\rm 1,thin}$ vs $\gamma_{\rm 1,thick}$. Table~\ref{tab_vel} in Appendix~\ref{MWdiskvel} gives our derived $v_{\rm\phi,thin}(R)$ and $v_{\rm\phi,thick}(R),$ shown in Fig.~\ref{MW_velprofiles}. 

\begin{table}
\caption{Best-fit parameters for the MW's rotational velocity profiles.}
\label{tab_fit_vphi}
\centering
\begin{tabular}{c c c}
\hline\hline
Component & thin disk & thick disk\\
\hline
$R_{\rm break}$ [kpc] & 5.47 $\rm\pm$ 0.03 & 4.56 $\rm\pm$ 0.06\\
$\gamma_{\rm 0}$ [km~s$^{\rm -1}$kpc$^{\rm -1}$] & 38.73 $\rm\pm$ 0.66 & 40.43 $\rm\pm$ 1.05\\
$v_{\rm 1}$ [km~s$^{\rm -1}$] & 237.30 $\rm\pm$ 0.65 & 204.86 $\rm\pm$ 2.41\\
$\gamma_{\rm 1}$ [km~s$^{\rm -1}$kpc$^{\rm -1}$] & -1.98 $\rm\pm$ 0.06 & -2.08 $\rm\pm$ 0.29\\
\hline
\end{tabular}
\end{table}

    
\subsection{AM of the MW stellar disks and DM halo}

For the MW's mass distribution by components, we applied the recent model from 
\citet{Cautun:2020}. These authors consider the following baryonic components: a bulge, a thin and a thick stellar disk, a gas disk (HI + molecular), and a gaseous halo, each with its own functional form. In their model, all the bulge parameters except total mass, the exponential scale
heights of the two stellar disks, and all the gas disk parameters are fixed to previously derived values \citep{2008ApJ...673..864J,2017MNRAS.465...76M}. The mass profile of their gaseous halo is derived from simulations \citep{Schaye:2015,Sawala:2016,Grand:2017,2019MNRAS.490.4786G}, such that it has no free parameters, and the total mass in this component is just a function of the total virial mass (dark matter + baryons). Therefore, the free parameters for the baryonic components in their model remain: total thin disk mass $M_{\rm thin}$ and scale length, $R_{\rm thin}$, total thick disk mass, $M_{\rm thick}$ and scale length, $R_{\rm thick}$, and total bulge mass, $M_{\rm bulge}$\footnote{The actual free parameters used by \citet{Cautun:2020} are the surface mass density normalizations $\Sigma_{\rm 0,thin}$ and $\Sigma_{\rm 0,thick}$, and the volume mass density normalization $\rho_{\rm 0}$ for the bulge.}. For the DM distribution, \citet{Cautun:2020} considered two possibilities: an uncontracted and a contracted Navarro-Frenk-White (NFW) profile \citep{Navarro:1997}. Their Bayesian model is constrained by the MW's rotation curve of \citet{Eilers:2019}, the virial mass estimate of \citet{2019MNRAS.484.5453C}, the ratio between thin and thick disks densities at $R_{\rm\odot}$ from the \citet{2018A&A...615L..15G}, and the vertical force at $(R,z)=(R_{\rm\odot},1.1 \mathrm{kpc})$ of \citet{1991ApJ...367L...9K}, and slightly favors the contracted NFW profile. They discuss the predictions of their two models with different DM profiles to other observables such as total stellar disk mass, escape velocity at $R_{\rm\odot}$, and stellar-to-DM mass ratio, concluding that these extra constrains also favor the contracted profile, but not definitively. 

To compute the angular momentum for the thin and thick disks, we use both of the models from \citet{Cautun:2020} along with the best-fit parameters from their Table 2. We symmetrize their reported uncertainties, such that we can use a multivariate normal distribution with a covariance that approximates the most relevant correlations in their Fig. 12 to propagate the uncertainties of their mass models into our estimates for the disks' AM. Table~\ref{mass_model_param}\footnote{The uncertainty in $J_{\rm h}$ (as well as in $j_{\rm h}$ and $\lambda_{\rm MW}$ implicitly) takes into account only the covariance and not the scatter of the $J_{\rm h}=f(J_{\rm disk})$ relation. Considering the 0.29~dex scatter of this relation, the values for $J_{\rm h}$, $j_{\rm h}$ and $\lambda_{\rm MW}$ should be considered certain within a factor of $\sim$2.} 
summarizes the parameters we adopted from \citet{Cautun:2020} and the assumed covariances are given in Appendix~\ref{model_mass}. 
In practice, we generated 10000 samples from the three independent covariance matrices of the AM, mass and the velocity models involved in Eqs.~\ref{eq_J}, \ref{eq_Jz_approx}, and \ref{eq_vel}. To compute the AM of the DM halo, we used Eq.~\ref{eq_J} with the best-fit parameters for combined disks in Table~\ref{table:2}. Figure~\ref{AM_distributions}
shows the resulting probability distributions functions (PDFs) of AM for the thin and thick disks (left), combined stellar disk (center), and DM halo (right) as thin black curves in each panel. The colored curves in the figure are the normal (left and central panel) and log-normal (right panel) distributions generated from the first and second order moment of the corresponding PDFs.
The larger masses for the thin and thick disks in the uncontracted NFW model 
naturally lead to larger estimates for the stellar AM of the disks, 
3.72$\rm\pm$0.49 vs 3.26$\rm\pm$0.43 $\rm\times$10$^{\rm 13}$M$_{\rm\odot}$kpc$\rm\cdot$km$\rm\cdot$s$^{\rm -1}$ for the thin disk and 1.43$\rm\pm$0.34 vs 1.20$\rm\pm$0.30 $\rm\times$10$^{\rm 13}$M$_{\rm\odot}$kpc$\rm\cdot$km$\rm\cdot$s$^{\rm -1}$ for the thick disk, and propagate into a larger AM of the DM halo, namely: 2.96$^{\rm +0.43}_{\rm -0.37}$ versus 2.69$^{\rm +0.37}_{-0.32}$ $\rm\times$10$^{\rm 15}$M$_{\rm\odot}$kpc$\rm\cdot$km$\rm\cdot$s$^{\rm -1}$. Table~\ref{AM_estimates} gives the AM and specific AM for the stellar disk, its two components, and the DM halo.

To place the MW in the broader context of nearby galaxies, we can look to, for instance, the Spitzer Photometry \& Accurate Rotation Curves sample \citep[SPARC, ][]{2016AJ....152..157L}, which is a compilation of 175 galaxies with good quality
rotation curves derived from previously published HI and H$_{\rm \alpha}$ data and with good photometry in the Spitzer 3.6 $\rm\mu$m band. In particular, \citet{2018A&A...612L...6P} used 92 galaxies from SPARC (with large enough radial coverage of the velocity profile and inclinations $>30^{\rm o}$) to argue that the specific AM of (stellar) disks is an unbroken power-law of disk mass over more than four orders of magnitudes in mass, from dwarfs to massive spirals. Interestingly, \citet{2018A&A...612L...6P} find a relatively small scatter for this relation, of only 0.15~dex. Placing the Galaxy in the $j_{\rm disk}$ -- $M_{\rm disk}$ plane, where $M_{\rm disk}=M_{\rm thin}+M_{\rm thick}$, we find it to be 1.7$\rm\sigma$ below the relation of Posti et al. for the uncontracted NFW model, and 1.1$\rm\sigma$ below for the contracted one. Therefore, if we take the relation of Posti et al. at face value, the contracted NFW model is favored over the uncontracted one.

\begin{table}
\caption{AM of MW's stellar disks and DM halo. The highlighted rows indicate properties whose values depend on the relation between the AM of disks and DM halos from simulations. The rest of the values are derived directly from the observational data alone.}
\label{AM_estimates}
\centering
\begin{tabular}{c c c}
\hline\hline
Property & contracted NFW & NFW\\
\hline
$J_{\rm z,thin}$ [10$^{\rm 13}$M$_{\rm\odot}$kpc$\rm\cdot$km$\rm\cdot$s$^{\rm -1}$] & 3.26 $\rm\pm$ 0.43 & 3.72 $\rm\pm$ 0.49\\
$J_{\rm z,thick}$ [10$^{\rm 13}$M$_{\rm\odot}$kpc$\rm\cdot$km$\rm\cdot$s$^{\rm -1}$] & 1.20 $\rm\pm$ 0.30 & 1.43 $\rm\pm$ 0.34\\
$J_{\rm z,disk}$ [10$^{\rm 13}$M$_{\rm\odot}$kpc$\rm\cdot$km$\rm\cdot$s$^{\rm -1}$] & 4.46 $\rm\pm$ 0.64 & 5.15 $\rm\pm$ 0.72\\[0.1cm]
\rowcolor{lightgray} $J_{\rm h}$ [10$^{\rm 15}$M$_{\rm\odot}$kpc$\rm\cdot$km$\rm\cdot$s$^{\rm -1}$] & 2.69 $^{\rm +0.37}_{-0.32}$ & 2.96 $^{\rm +0.43}_{\rm -0.37}$\\[0.1cm]

$j_{\rm z,thin}$ [kpc$\rm\cdot$km$\rm\cdot$s$^{\rm -1}$] & 1026 $\rm\pm$ 59 & 935 $\rm\pm$ 51\\
$j_{\rm z,thick}$ [kpc$\rm\cdot$km$\rm\cdot$s$^{\rm -1}$] & 1308 $\rm\pm$ 242 & 1343 $\rm\pm$ 216\\
$j_{\rm z,disk}$ [kpc$\rm\cdot$km$\rm\cdot$s$^{\rm -1}$] & 1089 $\rm\pm$ 72 & 1021 $\rm\pm$ 62\\[0.1cm]
\rowcolor{lightgray} $j_{\rm h}$ [kpc$\rm\cdot$km$\rm\cdot$s$^{\rm -1}$] & 2769 $^{\rm +589}_{-485}$ & 3617 $^{\rm +643}_{\rm -546}$\\[0.1cm]

\rowcolor{lightgray} $\lambda_{\rm MW}$ & 0.061 $^{\rm +0.022}_{\rm -0.016}$ & 0.088 $^{\rm +0.024}_{\rm -0.020}$\\[0.1cm]
\hline
\end{tabular}
\end{table}

    \begin{figure}
   \centering
   \includegraphics[width=0.5\textwidth]{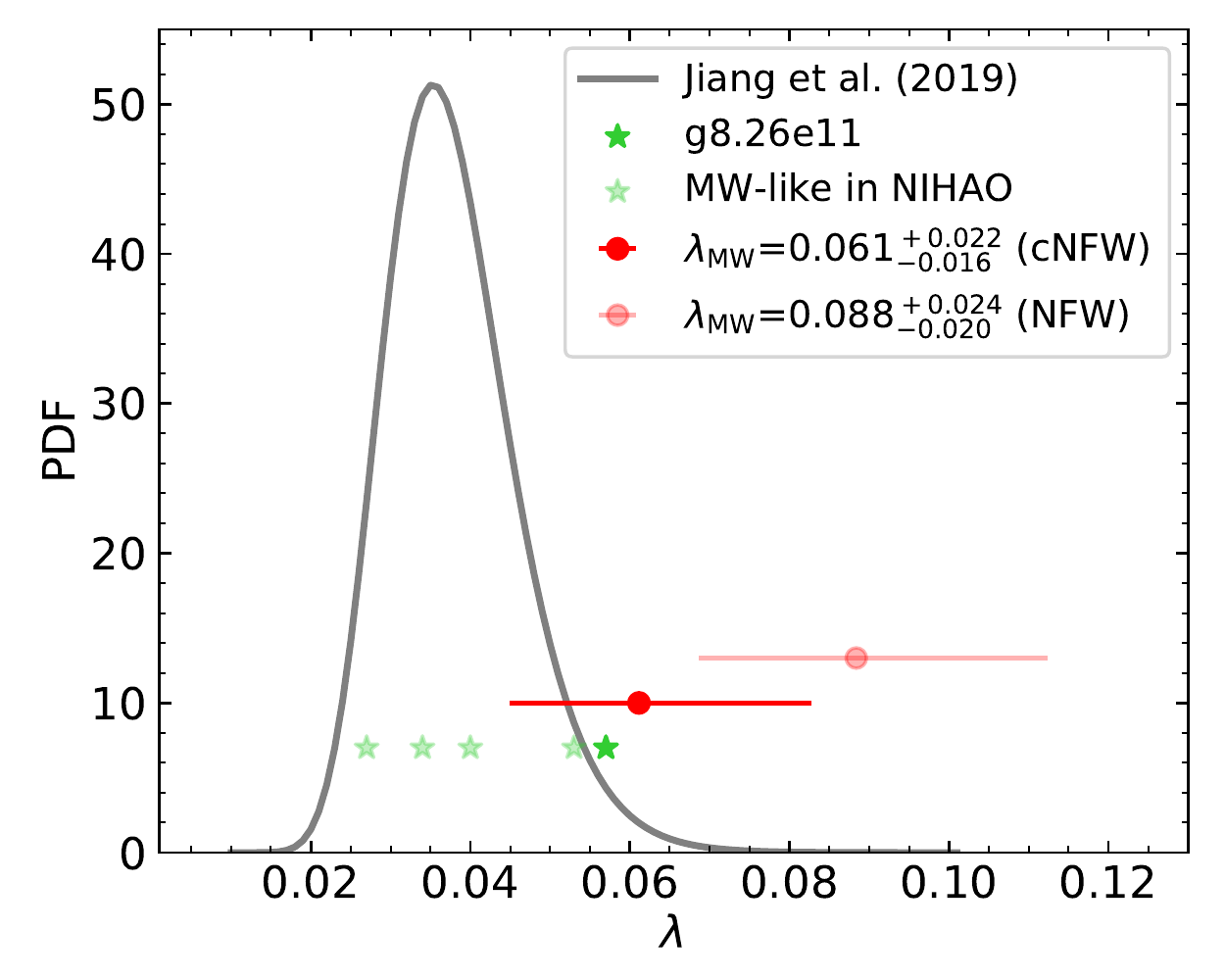}
   \caption{DM halo spin distribution for the NIHAO sample \citep[gray curve,][]{Jiang:2019}, together with the MW predictions for the NFW and contracted NFW mass models (faint and solid red points). The green stars mark the spins of the MW-type galaxies in NIHAO, among which galaxy g8.26e11 is the closest analog, both in terms of stellar and DM mass, as well as in terms of stellar disk structure \citep{Obreja:2018,2020MNRAS.491.3461B}. Most of the uncertainty in $\lambda_{\rm MW}$ comes from the uncertainties in the mass models of \citet{Cautun:2020}. The ordinates of the red and green points have no physical meaning.} 
   \label{MW_spin}%
    \end{figure}

To place the MW's DM halo in context, we computed the spin following the definition of \citet{Bullock:2001}:
\begin{equation}
 \lambda = \frac{J_h/M_h}{\sqrt{2}r_{\rm 200}v_{\rm 200}},
 \label{spin_eq}
\end{equation}
where the virial velocity is $v_{\rm 200}=\sqrt{G M_{\rm 200}/r_{\rm 200}}$, $M_{\rm 200}$ is the total virial mass, and $G$ is the gravitational constant. To compute the virial radius $r_{\rm 200}$ as the radius enclosing an average overdensity 200 times the critical matter density, we assume the following cosmological parameters from \citet{2020A&A...641A...1P}: $h = 0.6737$, $\Omega_{\rm m}= 0.3147$ and $\Omega_{\rm \Lambda} = 0.6853$, and the total halo mass $M_{\rm 200}$
of the MW corresponding to the contracted and uncontracted NFW profiles.

The values of the MW's halo spin for the two DM mass models are  reported in Table~\ref{AM_estimates}, and are shown in Fig.~\ref{MW_spin} together with the universal log normal distribution as quantified by \citet{Jiang:2019}: $\overline{\lambda}=0.037$
and $\sigma_{\rm ln\lambda}=0.215$. The large uncertainty in $\lambda_{\rm MW}$ is mostly driven by the uncertainties in the mass models of \citet{Cautun:2020}: reducing all the uncertainties in the mass models by a factor of 10 result in an uncertainty that is about four times lower.\footnote{Forcing the fit to the velocity profiles to  $R_{\break}=$~5~kpc and $v_{\rm 0}=$~0~km~s$^{\rm -1}$ changes the values in table~\ref{AM_estimates} by $\sim$1\% or less.}

The contracted NFW model results in a spin of $\lambda_{\rm MW}= 0.061^{\rm +0.022}_{\rm -0.016}$, which is $2.3\sigma_{\rm ln\lambda}$ away from the peak of the log normal distribution. In other words, the probability for a halo to have this value of $\lambda$ (within its uncertainty) is 21\%. For the uncontracted NFW model, the resulting spin $0.088^{\rm +0.024}_{\rm -0.020}$ is a more extreme outlier, $4.0\sigma_{\rm ln\lambda}$ away from the peak, which translates into a probability of only 0.2\%. This means that under our assumption that the AM of the DM halos can be derived from the AM of their stellar disks, the contracted NFW model for the MW's DM halo is strongly favored over the uncontracted one. In any case, both estimates of $\lambda_{\rm MW}$ imply that MW's DM halo spins significantly faster than what is expected for halos of similar masses.

To get an idea of how reasonable our derived value of $\lambda_{\rm MW}= 0.061$ is, 
we also show in Fig.~\ref{MW_spin} the spins of MW-type galaxies in NIHAO (green stars). From left to right, the five green stars represent the simulated galaxies: g7.66e11 ($\lambda=0.027$), g6.96e11 ($\lambda=0.034$), g7.08e11 ($\lambda=0.040$) g7.55e11 ($\lambda=0.053$) and g8.26e11 ($\lambda=0.057$). Figure~\ref{vc_profiles} in Appendix~\ref{MWs_prop} shows the circular velocity curves of these simulated galaxies in comparison with MW data. The last simulated galaxy, g8.26e11, which has a spin parameter very close to the one inferred for the Galaxy has already been shown to be a very good MW analog in terms of DM halo mass, total stellar mass, and structure of the thin and thick stellar disks (Paper I, \citealt{2020MNRAS.491.3461B}). This simulated galaxy had its last important merger $\sim$10~Gyr ago. The DM mass ratio of this event is 5.7:1, and this merger is responsible for $\sim60\%$ of the $z=0$ DM halo's AM (Fig. 8 in Paper I). Both the timing and the mass ratio of this merger are close to the values inferred for the event that is thought to have triggered Gaia Sausage \citep[major merger 8 to 10 Gyr ago,][]{Belokurov:2018} also known as Gaia Enceladus \citep[4:1 merger ratio $\sim$10~Gyr ago,][]{Helmi:2018}.  

The uncanny resemblance of g8.26e11 to MW raises the possibility that $\lambda$ should also be one of the important parameters (together with the $z=0$ halo mass and its mass accretion history) when selecting halos from DM-only simulations to re-simulate at higher resolution; the goal here is that they would  end up hosting MW analogs that are as good a fit as possible. For instance, if we consider the stellar mass function $\Phi(M_{\rm *})$ as computed from the combination of Galaxy And Mass Assembly survey \citep[GAMA, ][]{Wright:2017} and \citet{Bernardi:2013}, with a lower limit of 10$^{\rm 8}$M$_{\rm\odot}$, we get a 1\% probability\footnote{$p(M_{\rm MW})=\int_{M_{\rm MW,low}}^{M_{\rm MW,up}}\Phi(M_{\rm *})dM_{\rm *}/\int_{\rm 10^{\rm 8}M_{\rm\odot}}^{\rm \infty}\Phi(M_{\rm *})dM_{\rm *}$} of picking up a galaxy with the MW's stellar mass (5.04$^{\rm+0.43}_{\rm-0.52}\rm\times$10$^{\rm 10}$M$_{\rm\odot}$, as in the cNFW model of Cautun et al.). Given that $\lambda$ does not depend on $M_{\rm *}$, in a survey (or in a large scale hydrodynamical simulation) complete down to stellar masses of 10$^{\rm 8}$M$_{\rm\odot}$, the probability of finding a galaxy with the MW's stellar mass and the MW's spin would be 0.22\%\footnote{$p(M_{\rm *},\lambda)=p(M_{\rm *})p(\lambda)$}.
The results of this study are based on a limited number of galaxies, but in combination with other high-resolution simulations of MW analogs from the literature \citep[e.g.,][]{Sawala:2016,Grand:2017,2019MNRAS.490.4786G,2018MNRAS.481.4133G,2020MNRAS.498.2968L,2021ApJ...906...96A} the sample size could be increased. At the same time, comparing results across different physics implementations can gauge the detailed impact of baryonic physics.

The MW is known as a rather atypical spiral galaxy also because of its two largest satellites: the Large and Small Magellanic Clouds (LMC and SMC). Configurations of the type MW-LMC-SMC are thought to be quite rare  \citep[e.g.,][]{2011MNRAS.414.1560B,2013ApJ...770...96G,2021MNRAS.500.3776W}, but the probability of a LMC-mass satellite is higher in MW-M31 galaxy pairs \citep[e.g.,][]{Santos-Santos:2021}. The exact orbital history of these two satellites is still under debate, but currently it has been well established that LMC and SMC are on prograde orbits, close to the MW's plane \citep[e.g.][]{2016ARA&A..54..363D}. Irrespective of the detailed infall scenarios for LMC and SMC, it is very likely that these mergers brought in a significant amount of AM, making the MW's DM halo a highly spinning structure.
Finally, having an estimate of the DM halo's AM for the MW opens up the possibility of including the dark halo rotation in dynamical models of our Galaxy.  


\section{Conclusions}

Using a subsample of simulated galaxies from the NIHAO project, we show that 
the total dark matter halo angular momentum, $J_{\rm h}$, is tightly correlated with the 
azimuthal angular momentum, $J_{\rm z}$, of dynamical disks and stellar halos. 
We use the data driven definition of dynamical stellar structures from \citet{Obreja:2018,Obreja:2019}, where Gaussian mixture models are applied in the parameter space of normalized 
angular momentum -- normalized binding energy. 

The tightest correlation is with the dynamical stellar halos, which is not unsurprising 
given that dark matter and stellar halos share similar spatial extensions.  
The other important correlation we find is between $J_{\rm h}$ and the $J_{\rm z}$
of dynamical stellar disks, albeit with a larger scatter and a smaller 
correlation coefficient than for stellar halos ($rms=0.29$~dex and $r_{\rm P}=0.90$ vs
$rms=0.13$~dex and $r_{\rm P}=0.97$).  
This relation can be explained by the fact that the angular momentum evolution of the 
Lagrangian masses forming the $z=0$ stellar disks and their host dark matter halos 
are relatively similar. The larger scatter in this relation can be at least partially 
explained by angular momentum redistribution suffered by the progenitor gas of $z=0$ stellar disks as an effect of stellar feedback. All parameters of the power laws between $J_{\rm h}$ and $J_{\rm z}$ are given in Table~\ref{table:2}.

We 
think the reasons why we do find tight relations between the dark matter halo and stellar components 
angular momenta, while many other previous studies did not, are twofold: i) our data-driven definition of stellar dynamical structures is robust and realistic, and ii) we look at the physical 
property that is known to be conserved (at least to some degree), namely, the angular momentum -- and not at the specific angular momentum or spin parameter. Disks are expected to conserve a large part of their initial or maximum angular momentum, similar to the case of dark matter halos, and in contrast to bulges. Therefore, if we have a good definition of what a "disk" is, as in which stellar particles and stars belong to the  "dynamical"\ disk in a simulated or a real galaxy, we can more accurately test how closely disks follow the angular evolution of dark matter halos. And if such disks do follow the dark matter, we expect to find tighter correlations between the angular momenta of the two than in cases without a robust and physically meaningful definition for a disk. Our second reason is entirely related to the scatter in the relations between: i) halo mass and  stellar mass \citep[inverted $M_{\rm star}$--$M_{\rm h}$ relations, e.g.,][]{Moster:2018}; ii) disk scale length and stellar mass \citep[e.g.,][]{2016AJ....152..157L}; and iii) disk-to-total ratio and total stellar mass \cite[e.g.,][]{2014ApJ...788...11L}. The scatter in these three relations hinders any correlation between the spin or the specific angular momentum of dark matter halos and galaxy properties that can be derived from observations, such as stellar mass, disk mass, disk scale length, disk specific angular momentum, or stellar and baryonic spin.

In particular, the relation between $J_{\rm h}$ and $J_{\rm z,disk}$  can be  
applied in a straightforward manner with current observational data to derive angular momenta of dark matter halos hosting 
real galaxies, and, subsequently, to compute their spin parameters $\lambda$. 
We carried out such an exercise for the Milky Way using observational data from APOGEE and Gaia DR2,
with improved parallaxes from \citet{Hogg:2019}, the circular velocity curve estimate of 
\citet{Eilers:2019}, and the Galaxy mass model of \citet{Cautun:2020}. 

In the parameter space of circularities, namely, $\alpha$-enhancement, $j_{\rm z}/j_{\rm c}$--[O/Fe], 
the sample of stars from \citet{Hogg:2019} occupies two distinct but partially overlapping regions, which can be separated using clustering analysis (e.g., Gaussian mixtures) and subsequently associated 
with MW's thin and thick disks. This separation allows us to construct the rotational
velocity profiles for the two populations of stars, $v_{\rm\phi,thin}(R)$ and $v_{\rm\phi,thin}(R)$. For radii $R\gtrsim5$~kpc, we find both disks to be well described by linear relations $v_{\rm\phi}(R)=v_{\rm 1}+\gamma_{\rm 1}\cdot R$ with similar slopes $\gamma_{\rm 1}\simeq-2$~km~s$^{\rm -1}$~kpc$^{\rm -1}$, but different normalizations, that is, $v_{\rm 1,thin}=237.3\pm0.7$~km~s$^{\rm -1}$ and $v_{\rm 1,thick}=204.9\pm2.4$~km~s$^{\rm -1}$. These relations imply an almost constant velocity lag of the thick disk with respect to the thin one of $\sim$33~km~s$^{\rm -1}$ (measured at $R_{\rm\odot}$) for $R\gtrsim5$~kpc, and result in rotational velocities at the Sun's radius of $221.2\pm0.8$~km~s$^{\rm -1}$ and $188\pm3.4$~km~s$^{\rm -1}$ for the thin and thick disks, respectively. 

We calculated the angular momenta for the thin and the thick disks, $J_{\rm z,thin}$ and $J_{\rm z,thick}$ by convolving the derived $v_{\rm\phi,thin}(R)$ and $v_{\rm\phi,thin}(R)$ with the corresponding exponential mass profiles from \citet{Cautun:2020}. We used both MW mass models from \citet{Cautun:2020}, one assuming a contracted NFW profile for the dark matter halo and the other 
an uncontracted NFW. Thus, for the contracted and uncontracted NFW profiles, the two disks' angular momenta in units of 10$^{\rm 13}$M$_{\rm\odot}$kpc$\rm\cdot$km$\rm\cdot$s$^{\rm -1}$ are: 
 $J_{\rm z,thin}=$3.26$\rm\pm$0.43 / 3.72$\rm\pm$0.49 and $J_{\rm z,thick}=$1.20$\rm\pm$0.30 / 1.43$\rm\pm$0.34, and together they translate into an angular momentum for the dark matter halo of 
 $J_{\rm h}\rm = 2.69^{\rm +0.37}_{-0.32}/ 2.96^{\rm +0.43}_{\rm -0.37}$~$\rm\times$10$^{\rm 15}$M$_{\rm\odot}$kpc$\rm\cdot$km$\rm\cdot$s$^{\rm -1}$. Factoring in the virial masses of the models 
 for the contracted and uncontracted dark matter profiles, the spin parameter $\lambda_{\rm MW}$ is estimated to be $\rm0.061^{\rm +0.022}_{\rm -0.016}$ / $\rm0.088^{\rm +0.024}_{\rm -0.020}$, which is $2.3\sigma_{\rm ln\lambda}$ / $4.0\sigma_{\rm ln\lambda}$ away from the peak of the universal log normal spin distribution, as quantified by \citet{Jiang:2019}. In other words, the probability of a halo having a spin within the range derived for the NFW model is a factor of ten smaller than the probability corresponding to the $\lambda$ range of the contracted NFW (0.2\% vs 21\%). Therefore, from the estimate of $\lambda_{\rm MW}$, the contracted NFW model for the Galaxy is strongly favored over the uncontracted one. Comparing MW's spin estimate with spins of MW-type galaxies in NIHAO, we find that our best Galaxy analog (very similar halo and stellar masses, Fig.~\ref{vc_profiles} in the Appendix~\ref{MWs_prop}) also has a very similar parameter $\lambda=0.057$ to $\lambda_{\rm MW}$. Interestingly, this particular simulated galaxy experienced a major merger with similar characteristics (5.7:1 dark matter halo ratio $\sim$10~Gyr ago) as the one thought to be responsible for the Gaia Enceladus feature \citep[4:1 merger ratio $\sim$10~Gyr ago,][]{Helmi:2018}. Also, this merger event in the simulation is responsible for a significant fraction of the final angular momentum of this galaxy's dark matter halo.
 
Relations such as the ones we found between the dynamical 
stellar components of galaxies and their host dark matter halos open up the possibility
to compute the spin distribution for observed galaxies and compare it to the now well-established 
global distribution from simulations. In a more general sense, having estimates for the dark matter angular momenta of observed galaxies provides a unique new constraint on their formation histories and a closer glimpse into the properties of dark matter halos in the real universe.

\begin{acknowledgements}

    We thank the anonymous referee for their comments which helped improve this manuscript. We would also like to thank Fabrizio Arrigoni Battaia for the many discussions on this work, and to the Lorentz Center in Leiden for hosting the workshop {\tt Dynamical reconstruction of galaxies}, where many ideas pivotal to this work have been discussed.
    All figures have been made with \texttt{matplotlib} \citep{Hunter:2007}. We have also 
    used the Python libraries \texttt{numpy} \citep{Walt:2011}, \texttt{scipy} \citep{Jones:2001}, \texttt{scikit-learn} \citep{Pedregosa:2011}, \texttt{pandas} \citep{mckinney-proc-scipy-2010}, and \texttt{astropy} \citep{2013A&A...558A..33A,2018AJ....156..123A}. \texttt{F2PY} \citep{Peterson:2009} and \texttt{pynbody} \citep{Pontzen:2013} have been used in parts of the simulation analysis.
    This research was carried out on the High Performance Computing resources at New York University Abu Dhabi; on the \textsc{isaac} cluster of the Max-Planck-Institut f\"{u}r Astronomie and on the \textsc{hydra} clusters at the Rechenzentrum in Garching. We greatly appreciate the contributions of these computing allocations.
    AO is funded by the Deutsche Forschungsgemeinschaft (DFG, German Research Foundation) -- 443044596. TB acknowledges support from the European Research Council under ERC-CoG grant CRAGSMAN-646955. 
\end{acknowledgements}

\bibliographystyle{aa}
\bibliography{MWspin}

\begin{appendix}

\section{MW-like galaxies in NIHAO}
\label{MWs_prop}

\begin{figure}[t!]
\centering
\includegraphics[width=0.45\textwidth]{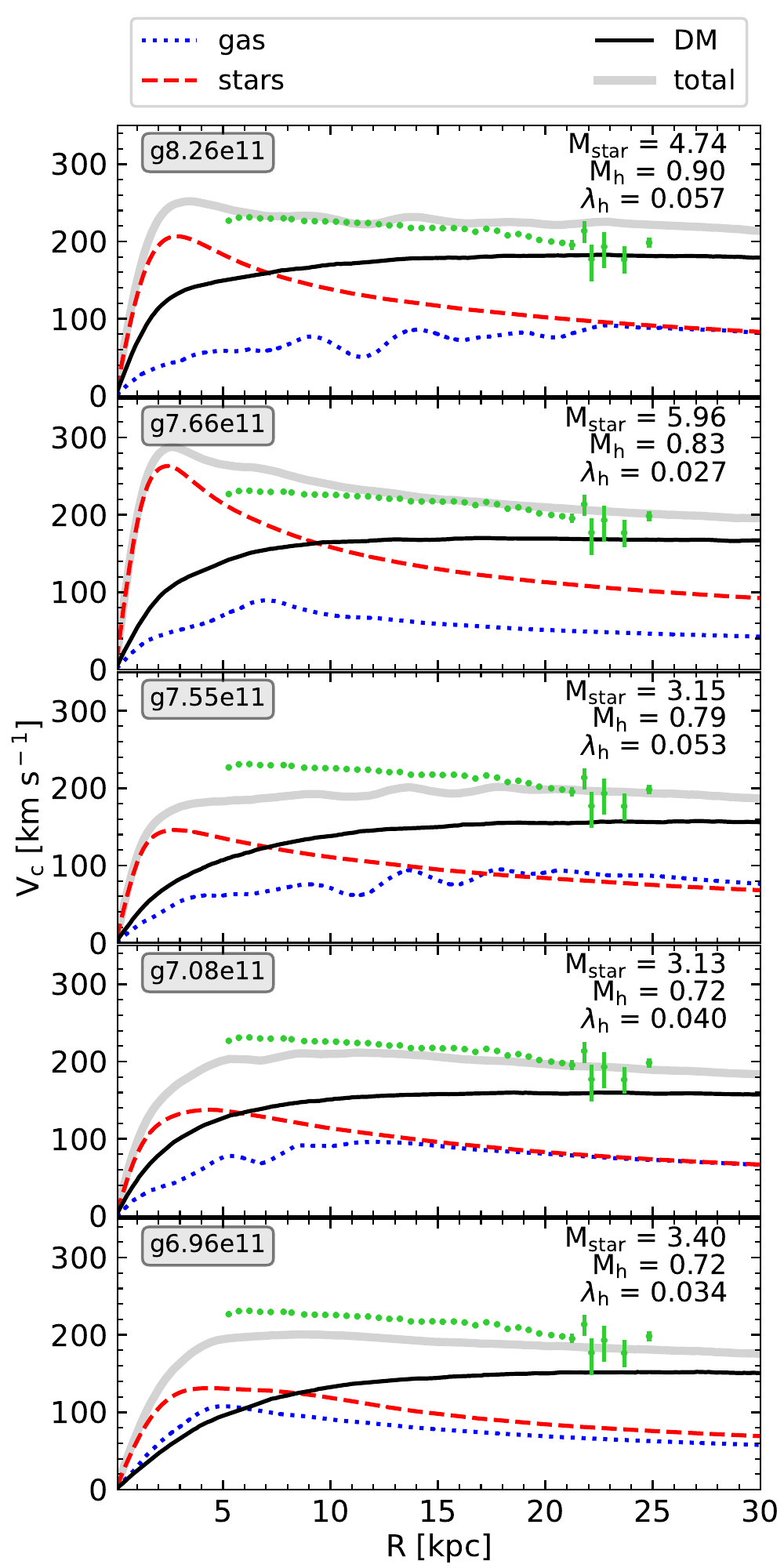}
\caption{Circular velocity profiles of MW-like galaxies in NIHAO. DM halo mass, $M_{\rm h}$, in units of 10$^{\rm 12}$M$_{\rm \odot}$, halo spin, and total stellar mass, $M_{\rm star}$, in units of 10$^{\rm 10}$M$_{\rm\odot}$ are quoted in the top right of each panel. MW observational data (green points) are from \citet{Eilers:2019}.} 
\label{vc_profiles}
\end{figure}

Figure~\ref{vc_profiles} shows the circular velocity profiles (thick gray lines) in the equatorial plane $V_{\rm c}=\sqrt{R \partial\Phi/ \partial R}$ for the five NIHAO galaxies that have stellar and DM masses within the ranges estimated for the MW. The galaxy with the most massive halo among the five, g8.26e11, shown in the top panel, has also the closest $\lambda=0.057$ to $\lambda_{\rm MW}=0.061$ for the contracted NFW model. This simulated galaxy has been shown to resamble MW in many aspects: i) thin- and thick-disk properties in terms of scale lengths and rotational velocity profiles \citep{Obreja:2018}; ii) vertical velocity dispersion vs stellar age and iii) thin- and thick-disk scale heights at $R_{\rm\odot}$ \citep{2020MNRAS.491.3461B}; iv) bimodal 2D distribution in [$\alpha$/Fe] vs [Fe/H] at $R_{\rm\odot}$ \citep{2020MNRAS.491.5435B}, and v) satellites distribution and properties \citep{Buck:2018a}. From the stellar and DM contributions to the $V_{\rm c}(R)$ of this galaxy (dashed red and solid black curves), it can be appreciated that DM becomes the dominant term at $R\simeq8$~kpc, similar to what \citet{Portail:2017} found from their dynamical model of MW, which includes the bar.    

The two main differences between g8.26e11 and our Galaxy is that the simulation has no bar or only a very weak bar, and it does not have satellites as massive as LMC and SMC
\citep{Buck:2018a}. 

\section{MW's rotational velocities}
\label{MWdiskvel}

Table~\ref{tab_vel} gives $v_{\rm\phi}(R)$ for the two disks of the Galaxy. These data have been used in Fig.~\ref{MW_velprofiles} and in the fit of Eq.~\ref{eq_vel}.

\begin{table}[hb!]
\caption{Rotational velocities profiles of MW's thin and thick disks.}
\label{tab_vel}
\centering
\begin{tabular}{c c c}
\hline\hline
$R$ [kpc] & $v_{\rm\phi,thin}$ [km s$^{\rm -1}$] & $v_{\rm\phi,thick}$ [km s$^{\rm -1}$]\\
\hline
5.25 & 218.85$\rm\pm$1.77 & 193.35$\rm\pm$2.92\\
5.75 & 223.81$\rm\pm$1.24 & 195.50$\rm\pm$3.10\\
6.25 & 223.60$\rm\pm$1.41 & 194.11$\rm\pm$3.16\\
6.75 & 221.72$\rm\pm$1.31 & 195.87$\rm\pm$3.00\\
7.25 & 221.01$\rm\pm$1.06 & 190.51$\rm\pm$2.78\\
7.75 & 221.60$\rm\pm$0.69 & 188.01$\rm\pm$1.84\\
8.25 & 221.62$\rm\pm$0.77 & 185.15$\rm\pm$2.21\\
8.75 & 218.64$\rm\pm$0.84 & 191.17$\rm\pm$3.32\\
9.25 & 219.30$\rm\pm$0.55 & 182.92$\rm\pm$2.84\\
9.75 & 217.18$\rm\pm$0.47 & 184.83$\rm\pm$2.52\\
10.25 & 216.93$\rm\pm$0.50 & 184.33$\rm\pm$2.90\\
10.75 & 216.34$\rm\pm$0.45 & 181.89$\rm\pm$2.78\\
11.25 & 215.44$\rm\pm$0.47 & 179.06$\rm\pm$3.51\\
11.75 & 215.63$\rm\pm$0.49 & 176.38$\rm\pm$3.16\\
12.25 & 214.46$\rm\pm$0.48 & 181.83$\rm\pm$4.55\\
12.75 & 213.06$\rm\pm$0.50 & 185.16$\rm\pm$4.14\\
13.25 & 211.58$\rm\pm$0.57 & -\\
13.75 & 209.79$\rm\pm$0.57 & -\\
14.25 & 207.37$\rm\pm$0.65 & -\\
14.75 & 206.79$\rm\pm$0.85 & -\\
15.25 & 204.59$\rm\pm$1.09 & -\\
15.75 & 205.78$\rm\pm$1.41 & -\\
16.25 & 203.89$\rm\pm$1.73 & -\\
16.75 & 201.61$\rm\pm$1.73 & -\\
17.25 & 203.03$\rm\pm$1.66 & -\\
17.75 & 202.81$\rm\pm$1.86 & -\\
18.25 & 202.26$\rm\pm$2.37 & -\\
18.75 & 193.71$\rm\pm$2.35 & -\\
19.25 & 196.02$\rm\pm$3.02 & -\\
19.75 & 191.09$\rm\pm$4.40 & -\\
\hline
\end{tabular}
\end{table}

\section{MW mass model}
\label{model_mass}

The free parameters of the MW mass model of \citet{Cautun:2020}, which we needed for 
computing the AM of the thin and thick disks, are given in Table~\ref{mass_model_param}. The uncertainties on the parameters have been symmetrized such that we can use a use a six-variate Gaussian distribution with a given covariance matrix to approximate the model of Cautun et al. The most important correlations in these models are among the DM halo mass $M_{\rm h}$, and thin and thick disk masses, $M_{\rm thin}$ and $M_{\rm thick}$, as can appreciated from their Fig. 12. 

For the contracted NFW model, we found the following covariance matrix:  
\begin{equation}
cov = 
\begin{pmatrix}
\sigma_1^2 & 0.7\sigma_1 \sigma_2 & 0.6\sigma_1 \sigma_3 & 0 & 0 & 0\\
0.7\sigma_1 \sigma_2 & \sigma_2^2 & 0.8\sigma_2 \sigma_3 & 0 & 0 & 0\\
0.6\sigma_1 \sigma_3 & 0.8\sigma_2 \sigma_3 & \sigma_3^2 & 0 & 0 & 0\\
0 & 0 & 0 & \sigma_4^2 & 0 & 0\\
0 & 0 & 0 & 0 & \sigma_5^2 & 0\\
0 & 0 & 0 & 0 & 0 & \sigma_6^2\\
\end{pmatrix}
\end{equation}
to reproduce well the 68 percentile contours in the panels $M_{\rm thin}$ versus log($M_{\rm h}$), 
$M_{\rm thick}$ versus log($M_{\rm h}$), and $M_{\rm thin}$ vs $M_{\rm thick}$ of Fig. 12 in
Cautun et al. We neglect the weak correlations with the other parameters. The corresponding indices 
of each parameter are listed in the last column of Table~\ref{mass_model_param}.

For the uncontracted NFW model, we use instead the following covariance matrix:
\begin{equation}
cov = 
\begin{pmatrix}
\sigma_1^2 & 0.7\sigma_1 \sigma_2 & 0.6\sigma_1 \sigma_3 & 0 & 0 & 0\\
0.7\sigma_1 \sigma_2 & \sigma_2^2 & 0.7\sigma_2 \sigma_3 & 0 & 0 & 0\\
0.6\sigma_1 \sigma_3 & 0.7\sigma_2 \sigma_3 & \sigma_3^2 & 0 & 0 & 0\\
0 & 0 & 0 & \sigma_4^2 & 0 & 0\\
0 & 0 & 0 & 0 & \sigma_5^2 & 0\\
0 & 0 & 0 & 0 & 0 & \sigma_6^2\\
\end{pmatrix}
\end{equation}

\begin{table}[ht!]
\caption{Parameters of the MW mass models of \citet{Cautun:2020}. Here, 
$M_{\rm gas}$ sums up the disk (HI + molecular) and the CGM, both contributions being kept fixed in the mass models. The bulge and gas masses are needed to compute the total viral mass $M_{\rm 200}$ in Eq.~\ref{spin_eq}.}
\label{mass_model_param}
\centering
\begin{tabular}{c c c c}
\hline\hline
Parameter & contracted NFW & NFW & index\\
\hline
log($M_{\rm h}$/M$_{\rm\odot}$) & 11.987 $\rm\pm$ 0.095 & 11.914 $\rm\pm$ 0.076 & 1\\
$M_{\rm thin}$ [10$^{\rm 10}$M$_{\rm\odot}$] & 3.18 $\rm\pm$ 0.38 & 3.98 $\rm\pm$ 0.47 & 2\\
$M_{\rm thick}$ [10$^{\rm 10}$M$_{\rm\odot}$] & 0.92 $\rm\pm$ 0.16 & 1.07 $\rm\pm$ 0.19 & 3\\
$R_{\rm thin}$ [kpc] & 2.63 $\rm\pm$ 0.13 & 2.43 $\rm\pm$ 0.11 & 4\\
$R_{\rm thick}$ [kpc] & 3.80 $\rm\pm$ 0.72 & 3.88 $\rm\pm$ 0.65 & 5\\
$M_{\rm bulge}$ [10$^{\rm 10}$M$_{\rm\odot}$] & 0.94 $\rm\pm$ 0.09 & 0.92 $\rm\pm$ 0.09 & 6\\
$M_{\rm gas}$ [10$^{\rm 10}$M$_{\rm\odot}$] & 7.6 & 6.7 & -\\
\hline
\end{tabular}
\end{table}

\end{appendix}

\end{document}